\documentstyle[12pt,epsfig]{article}

\newcommand{\bs}{\bar{s}}
\newcommand{\bu}{\bar{u}}
\newcommand{\bd}{\bar{d}}  
\newcommand{\amz}{$\alpha_s(M_Z^2)$}
\newcommand{\as}{$\alpha_s$}

\def\Title#1{\begin{center} {\Large {\bf #1} } \end{center}}

\begin{document}

\Title{Structure Functions  \\ in Deep Inelastic Lepton-Nucleon Scattering}

\bigskip\bigskip


\begin{raggedright} 
\begin{center} 
{\it Max Klein\index{Author, A.B.} \\
\vspace{0.3cm}
DESY/Zeuthen, Platanenallee 6, 
D-15738 Zeuthen ~~~~~~~~~ klein@ifh.de }
\end{center}
\bigskip\bigskip

Latest results on structure functions, as available at the 
Lepton-Photon Symposium 1999,
are presented.  This report focusses on three experimental areas:
new structure function measurements, in particular from HERA at 
low $x$ and high $Q^2$; results on light and heavy flavour densities;
determinations of the gluon distribution and of $\alpha_s$. As 
the talk was delivered at a historic moment and place,
a few remarks were added recalling the exciting past 
and looking into the promising future of
 deep inelastic scattering (DIS)\index{deep inelastic scattering}.
\end{raggedright}  
\section{Introduction}
About three decades ago,
 highly inelastic electron-proton scattering was observed by a
SLAC-MIT Collaboration~\cite{taylor} which measured the proton structure 
function\index{proton structure function}
 $\nu W_2(Q^2,\nu)$ to be independent of the four-momentum
transfer squared $Q^2$ at fixed Bjorken $x=Q^2/2M_p\nu$. Here
$\nu=E-E'$ is the energy transferred by the virtual photon. It is related 
to the inelasticity $y$ through $\nu=sy/2M_p$, with  proton mass $M_p$
and the energy squared in the centre of mass system  $s=2M_pE$.
 With the SLAC linear accelerator the incoming
electron energy $E$ had been successfully increased by a factor of twenty
as compared to previous form factor experiments~\cite{hofst}. Thus
$Q^2 = 4 E E' \sin^2(\theta/2)$ could be  enlarged
and measured using the scattered electron energy $E'$ and its polar
angle $\theta$. Partonic proton substructure~\cite{feyn}
was established at  $1/\sqrt{Q^2} \simeq 10^{-16}$~m 
which allowed the scaling  behaviour~\cite{bj}
of $\nu W_2(Q^2,\nu) \rightarrow F_2(x)$ to be interpreted.
 In the quark-parton model (QPM)~\cite{qpm}
the structure function $F_2$ is given by the 
momentum distributions of valence and sea quarks, $q=q_v+q_s$,
and of antiquarks $\bar{q}$ weighted by the square  $Q_q^2$ of the
 electric charge,
%
 $F_2(x,Q^2) = x \sum_q Q_q^2 ( q +\bar{q})$.
%
Neutrino experiments found $\sigma_{\nu} \simeq 3~\sigma_{\bar{\nu}}$
demonstrating that partons could be identified
with quarks having gauge couplings like leptons
 and  that at large $x$ the sea is small.
Scaling violations were hidden in the first DIS
data taken at $x \simeq 0.2$, as if we needed help to understand
the basics of inelastic scattering. They were
found in $\mu N$ scattering~\cite{loken} in an extended  $x, Q^2$
range.
The logarithmic $Q^2$ dependence of $F_2(x,Q^2)$,
established in subsequent neutrino and muon-nucleon scattering 
experiments, was attributed to quark-gluon interactions
in Quantum Chromodynamics~\cite{qcd}.
\begin{figure}[htb]
\begin{center}
\begin{picture}(200,140)
\put(-30,-20){
\epsfig{file=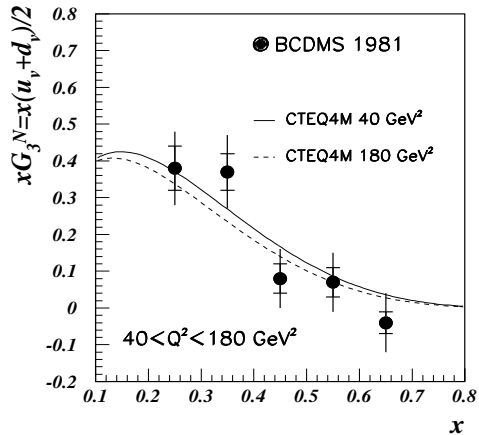,
width=12cm,height=16cm,bbllx=0pt,bblly=0pt,bburx=557pt,bbury=792pt}}
\end{picture}
\caption{Measurement of the $\gamma Z$ interference structure function
$xG_3$ in $\mu^{\pm}$~Carbon scattering by the BCDMS
 Collaboration compared with
a recent parametrization of the valence quark distributions by the
CTEQ group.}
\label{xg3}
\end{center}
\end{figure}
%

With the discovery of  neutral currents~\cite{ncrmp}
DIS neutrino experiments
made a major contribution to the theory of weak interactions.
In 1979 another $ep$ scattering experiment  
was performed at SLAC~\cite{presc}
which determined in a highly sensitive polarization
asymmetry measurement at $Q^2 \simeq 1.5$~GeV$^2$ the 
right-handed weak isospin charge of the electron
to be zero. This experiment 
selected thus the Glashow Weinberg Salam model
as the standard electroweak theory
and opened the possibility to
investigate proton structure at high $Q^2$ via $Z$ boson exchange.
The nucleon structure function $F_2$ was generalized, still in
a $V-A$ theory~\cite{derman},
to three functions
\begin{equation}
 (F_2, G_2, H_2) = x \sum_q (Q_q^2, 2Q_qv_q, v_q^2+a_q^2) (q+\bar{q})
\label{fgh}
\end{equation}
arising from photon exchange ($F_2$), $\gamma Z$ interference ($G_2$) and
 $Z$ exchange ($H_2$),
 where $v_q (a_q)$ are the vector
(axial vector) quark couplings~\cite{mkt}. 
In charged lepton-nucleon neutral current\index{neutral current} (NC)
scattering
two further structure functions appear
which are analogous to $xF_3$ in
neutrino scattering
\begin{equation}
 (xG_3, xH_3) = 2x \sum_q (Q_qa_q, v_qa_q) (q-\bar{q}).
\label{gh}
\end{equation}
A DIS muon experiment 
with simultaneous  beam charge and polarity reversal 
resulted in the first determination of the
$\gamma Z$ interference structure function $xG_3$
at $Q^2 \simeq 60$~GeV$^2$
by the BCDMS Collaboration at CERN, Fig.~\ref{xg3}.
 Electroweak interference occurs at the level of $\kappa
\simeq 10^{-4} Q^2 /$~GeV$^2$ as defined by
the ratio of the weak and the electromagnetic coupling constants.
Since the axial vector couplings could be considered to be known
this was an interesting measurement of the valence
 quark distribution sum $u_v + d_v$
which confirmed the sign of the quark charge combination
$Q_u-Q_d$ to be positive.

\begin{figure}[htb]
\begin{center}
\begin{picture}(200,175)
\put(-110,-25){
\epsfig{file=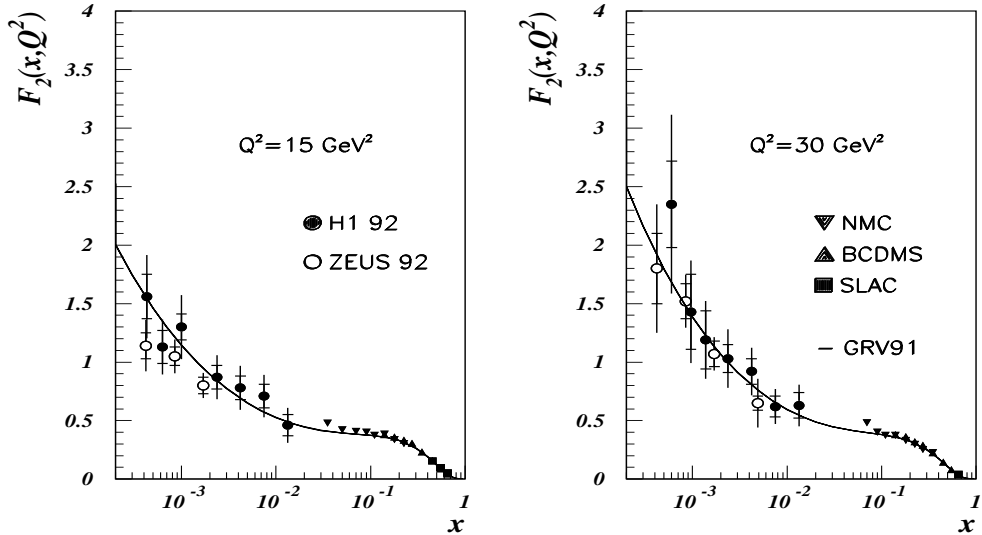,
width=14cm,height=16.5cm,bbllx=0pt,bblly=0pt,bburx=557pt,bbury=792pt}}
\end{picture}
\caption{Measurements of the proton structure function $F_2(x,Q^2)$
by the muon-proton scattering experiments BCDMS and NMC and their
extension towards low $x$ by the first available HERA data
on $F_2$. The curve
represents the anticipation of this rising behaviour 
by Gl\"uck, Reya and Vogt~\cite{grv91}.}
\label{92data}
\end{center}
\end{figure}

With
the HERA energy of $s=4E_eE_p \simeq 10^5$~GeV$^2$ the kinematic range
of DIS experiments could be
 greatly extended towards high $Q^2$ since $s$ was enlarged
 by a factor of about $2 E_p/$GeV
compared to fixed target scattering. The
first measurements of $F_2$ by the H1~\cite{h1f292} 
and the ZEUS~\cite{zeus92} Collaborations, 
using data taken in 1992, reached $x \simeq 0.0005$
at $Q^2 \simeq 20$~GeV$^2$. 
They discovered  a steep rise of
$F_2(x,Q^2)$ towards low $x$ at fixed $Q^2$:
below $x \simeq 0.01$ a decrease
by one order of magnitude translates into an increase of $F_2$
by about a factor of two, Fig.~\ref{92data}.
Although  a ``Possible Non-Regge Behaviour 
of Electroproduction Structure Functions''~\cite{heros} at low $x$
 had been considered and 
the concept~\cite{dynpart} and modified phenomenology~\cite{grv91} of
`dynamical partons' had been worked out,
 this rise came as some surprise since the DGLAP
evolution equations do not $a~priori$ fix the $x$ behaviour.
This rise
is now basically understood as being due to the dominance of gluons 
which 
leads to the description of the scaling violations
as $\partial{F_2}/\partial{\ln Q^2} \propto \alpha_s \cdot xg$
for $Q^2 \gg M_p^2$ and low $x$.
 Its quantitative
description in NLO QCD and the search for new dynamics~\cite{job}
 connected with
large logarithms of $1/x$ requires highest possible precision, i.e.
improved instrumentation and
higher luminosity than was available
when the first observation was made. 

While much attention has been paid to the inclusive and
charm structure function 
measurements at HERA, remarkable progress was also achieved 
in the investigation of
up, down, strange and charm quark distributions with neutrino and
Drell-Yan experiments at the Tevatron.

This paper describes a talk on structure functions
in deep-inelastic scattering 
delivered in 1999. Such a report is to some extent 
personal and cannot possibly cover 
 this expanding field of particle physics
in any exhaustive fashion.
 It thus may be seen together with further
articles, e.g.~\cite{marfort, alta}, and with the conference on 
deep inelastic scattering and QCD
held at Zeuthen in April 1999~\cite{DIS99}.
It demonstrates remarkable progress 
in DIS since the previous Symposium on Lepton-Photon 
Interactions~\cite{vs}. This talk focussed on recent
measurements of structure functions (Section~2), of quark distributions
including charm (Section~3) and determinations
of the gluon distribution and of $\alpha_s$ (Section~4). 
The field of deep inelastic lepton-nucleon
scattering has an  exciting future as will be
 described briefly in Section~5.
\section{Recent Measurements of Structure Functions}
\begin{figure}[htb]
\begin{center}
\begin{picture}(200,215)
\put(-95,-100){
\epsfig{file=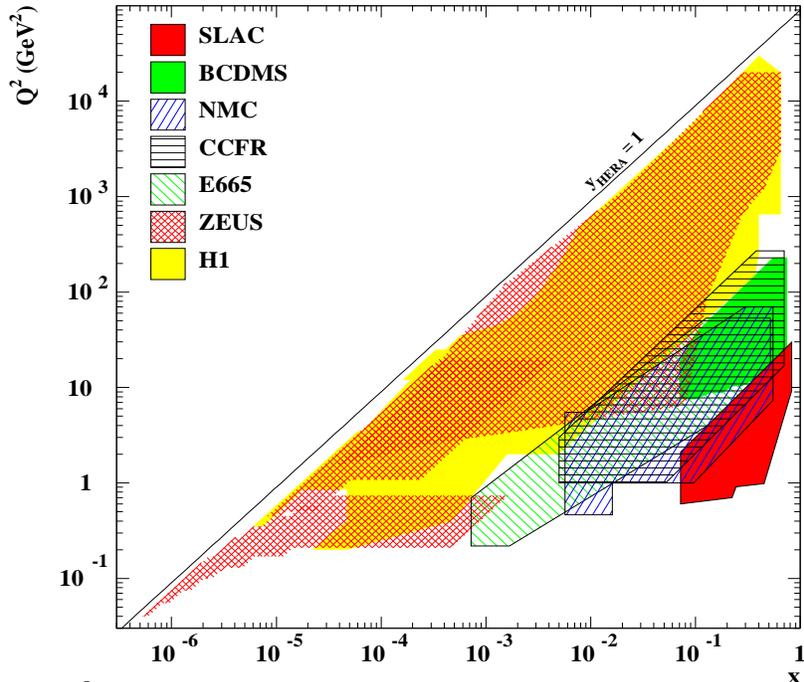,
width=11cm,height=14.7cm,bbllx=0pt,bblly=0pt,bburx=557pt,bbury=792pt}}
\end{picture}
\caption{$x, Q^2$ range covered by fixed target and collider
deep inelastic neutral current
 scattering experiments carried out until 1997.}
\label{kine}
\end{center}
\end{figure}

Since the first SLAC experiment,
fixed target muon and neutrino-nucleon scattering
experiments and subsequently the HERA collider experiments H1 and ZEUS
extended the explored kinematic region of DIS   
by several orders of magnitude, Fig.~\ref{kine}.
At smallest $x$  partons
carry only a vanishing fraction of the proton momentum. 
Hence the kinematics resembles 
the fixed target experiments where
both the electron and hadrons are scattered 
into the lepton beam direction (unfortunately termed 
`backward' at HERA). For high $Q^2 > s x E_e / (E_e + x E_p)$,
i.e. $Q^2 > 2,800$~GeV$^2$ for $x > 0.5$,
the electron is scattered through  angles $\theta > 90^o$ with respect to
the electron beam direction, similar to Rutherford  backscattering.
The kinematic range of the HERA collider experiments is confined
to about $y \geq 0.001$. For lower $y$  hadrons escape
in the forward (proton beam) direction. At very small
$y$ the inclusive kinematics cannot be
reliably reconstructed using the scattered electron variables alone
since the $x$ resolution varies like $1/y$.

Until 1997 HERA ran with positrons scattered off protons of
820~GeV energy and about 40~pb$^{-1}$ of luminosity became
available for each collider experiment.
From 1998 till May 1999  data samples 
of about 15~pb$^{-1}$ were collected in  collisions of electrons
with 920~GeV protons. The $e^{\pm}$ energy is tuned to about 27.5 GeV
to optimize the  polarization for the
fixed target experiment HERMES.
Longitudinal lepton beam polarization is foreseen to be used in
colliding beam mode from 2001 onwards.
%
%
\subsection{Transition to Photoproduction and 
Low \boldmath{$Q^2 \leq M_p^2$}}
\begin{figure}[htb]
\begin{center}
\begin{picture}(200,240)
\put(-60,-20){
\epsfig{file=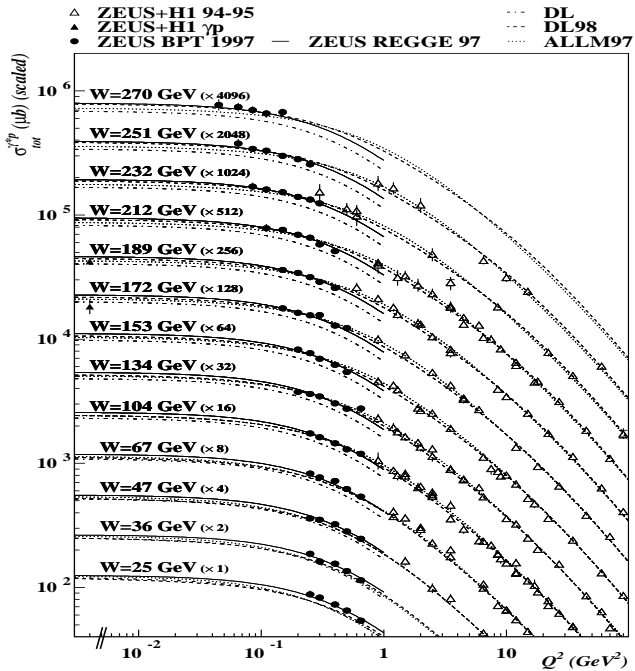,
width=10cm,height=10cm,bbllx=0pt,bblly=0pt,bburx=591pt,bbury=841pt}}
\end{picture}
\caption{
Measurements of the proton structure function $F_2(x,Q^2)$
expressed as $\sigma_{tot}^{\gamma^*p}(Q^2,W^2)$ from 
recent low $Q^2$ data by ZEUS (closed points) and
rebinned $F_2$ data at larger $Q^2$ by H1 and ZEUS (open squares).
The two triangles at
$W=172$~GeV and $W=189$~GeV near $Q^2 \approx 0$ denote the direct
photoproduction cross section measurements of H1 and ZEUS.}
\label{zeuslq}
\end{center}
\end{figure}

%
The structure function $F_2$ which dominates the DIS cross section
behaves like $x^{-\lambda(x,Q^2)}$ and vanishes  due 
to gauge invariance
 with $Q^2 \rightarrow 0$ like \cal{O}$(Q^2)$.
 The total virtual photon-proton scattering cross 
section is related to $F_2$ as
 $\sigma_{tot}^{\gamma^*p} \simeq  4 \pi^2 \alpha \cdot F_2 /Q^2$.
 Measurements of $F_2$ at low $Q^2$ investigate the dynamics of
 the transition from the deep inelastic to the photoproduction 
 regime~\cite{trans}. In Regge theory the structure function $F_2$
 results from a superposition of exchanged Regge poles with
 intercepts $\alpha_i$, $F_2 = \sum {\beta_i(Q^2) W^{2 \alpha_i -2}}$,
 where $W^2 \simeq Q^2/x \gg Q^2$ for low $x$, $W$ being the invariant
  mass of the $\gamma^*p$ system. A recent fit to $F_2$ data (DL98)
 is rather successful
  using three trajectories, i.e. $\alpha_1 = 1.08$ for the soft pomeron,
  $\alpha_2 = 0.55$ for $a$ and $f$ exchange 
  and $\alpha_3 = 1.4$ for 
  the so-called hard pomeron~\cite{dl98}. 
  For $Q^2 \rightarrow 0$ the exponent
 $\lambda$ is approximately
 given by the dominant pomeron Regge trajectory, i.e.
 $\lambda \simeq \alpha_1 -1 \simeq 0.1$. 
  The recent ZEUS data~\cite{zeusbpt},
  obtained with a backward calorimeter and tracker 
  positioned close to the beam pipe,
  are rather well described by this model, 
  Fig.~\ref{zeuslq}.
 Phenomenological models using 
  a combination of Generalized Vector Meson Dominance~\cite{gvdm}
   and perturbative QCD~\cite{allm}
 describe this transition also well.   
Extrapolations
  of  $F_2(x,W^2)$ to $Q^2 \simeq 0$ come out to be somewhat higher
  than the direct measurements 
   of  $\sigma_{tot}^{\gamma^*p}$~\cite{zeusbpt} with tagged electrons.
 The $F_2$ based 
  $\sigma_{tot}^{\gamma^*p}$ data are still at some $Q^2$ distance 
  from the real photoproduction measurements which have
 uncertainties of about 10\% due to beam optics and the imperfect
 simulation of the complete final state.  Further
extension of the range of the inclusive $F_2$ measurements at HERA
 towards lowest $Q^2$ values is thus desirable.
   This could be achieved in a rather short run of HERA
at minimum possible electron beam energy since 
$Q^2$ is proportional to $E^2$ for all except the high $y$ values.
  
 New data on parton-hadron duality~\cite{bg} became available this year
 from an experiment at Jefferson Laboratory~\cite{cynthi}
 measuring electron-proton and deuteron 
 scattering in the resonance region $W \simeq 1$~GeV. The superposition
 of cross sections,
  determined at  different $Q^2$ between 0.2 and 3.3~GeV$^2$,
 leads to an averaged behaviour of $F_2$ which is valence like even 
 at low $x$, or mass corrected $\xi$~\cite{poli},
  which supports the assumption
 made in the GRV analysis~\cite{grv91}
   for the initial $x$ distributions at very 
 small $Q^2$. In this experiment, which in the future will measure the 
 ratio $R=\sigma_L/\sigma_T$, one estimates power corrections
 (`higher twists') to be small
 and derives the magnetic elastic proton form factor $G_M^p$ from
 inelastic data.   
\subsection{Neutrino Experiments}
\label{nuexp}
The final measurement of $\nu Fe$ and $\bar{\nu} Fe$ scattering
cross sections  by the CCFR 
Collaboration~\cite{ccfrsig} is in good agreement with
previous data obtained by the CDHSW Collaboration and more accurate.
The high statistics CCFR  data has been used for a number of 
investigations regarding all structure functions 
involved (Sections~\ref{H1prec},~\ref{FL})
and also for tests of QCD (Section~\ref{alphas}). Recently data were
released for extremely large $x > 0.75$ pointing to cumulative 
effects beyond Fermi motion in the nucleus~\cite{ccfrhix} which
were  studied previously  by the BCDMS Collaboration~\cite{bcdmshix}.

Data were obtained by the IHEP-JINR neutrino experiment 
in the wide band  neutrino beam
at the Serpukhov U70 accelerator~\cite{vovenko}. Based on
about 750 $\nu$ and 6000 $\bar{\nu}$ events for 
$W^2 > 1.7$~GeV$^2$ and $Q^2 \simeq 2$~GeV$^2$, the structure
functions $F_2$ and $xF_3$ were disentangled  and
$\alpha_s(M_Z^2) = 0.123^{+0.010}_{-0.013}$ was determined in NLO QCD.

\subsection{Precision Measurement at Low \boldmath{$x$}
 and Medium \boldmath{$Q^2$}}
\label{H1prec}
The H1 Collaboration released for this conference the so far
most precise measurement of the DIS cross section  at HERA. 
In its reduced form it can be written as 
\begin{equation}
 \frac{Q^4 x} {2 \pi \alpha^2 Y_+} \cdot 
 \frac{d^2\sigma}{dQ^2dx} = \sigma_r = 
       F_2 - \frac{y^2}{Y_+} \cdot F_L,
        \label{sigr}
\end{equation}
i.e. 
$\sigma_r \simeq F_2$ apart from high $y$ where
$\sigma_r \rightarrow F_2-F_L \propto \sigma_T$.
 Here $F_L$ denotes the longitudinal
structure function which is related to the ratio
$R=F_L/(F_2-F_L)$ and $Y_+=1+(1-y)^2$.
 The H1 data, taken in 1996 and 1997, have statistical
errors of typically 1\% 
and systematic errors of 2-3\%, apart from edges of the acceptance region. 
In order to reach this precision  HERA has been anually increasing
the luminosity. The H1 experiment was subject to 
a major upgrade of its backward
apparatus  replacing  a Pb-Scintillator calorimeter 
%
\begin{figure}[htb]
\begin{center}
\begin{picture}(200,320)
\put(-110,-30){
\epsfig{file=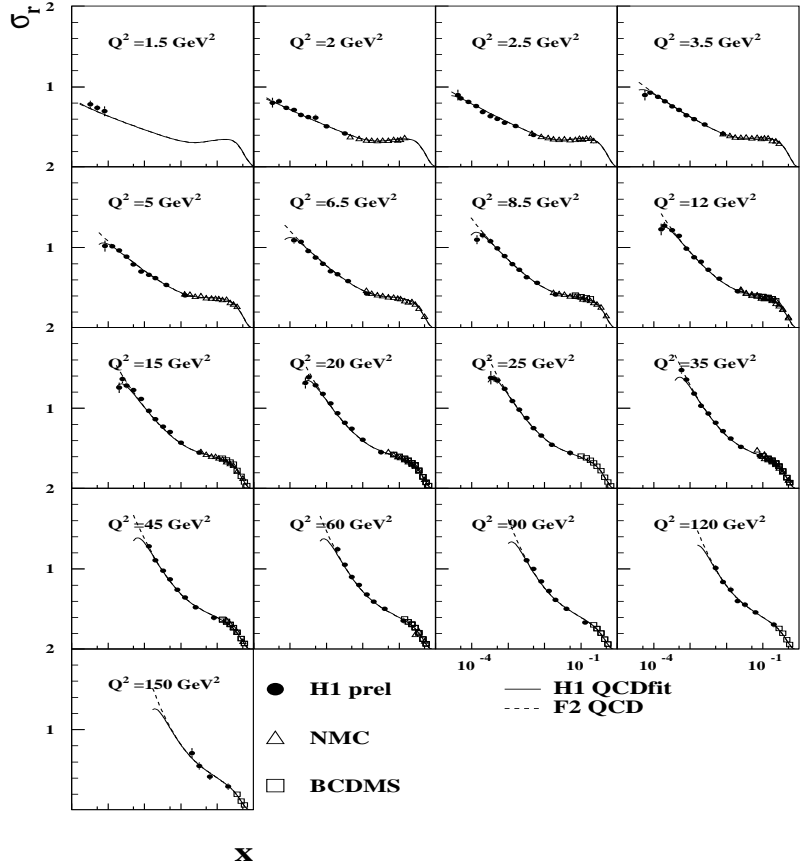,
width=12cm,height=12.5cm,bbllx=0pt,bblly=0pt,bburx=557pt,bbury=792pt}}
\end{picture}
\caption{
Measurements of the DIS cross section by H1 compared with 
NMC and BCDMS $\mu p$ data. The 
solid curve denotes the fitted cross section in NLO QCD
using H1 and NMC data for $Q^2 \geq 3.5$~GeV$^2$. The dashed curve
is the structure function $F_2$ obtained in the QCD fit
which at low $x$ departs from $\sigma_r$.}
\label{h1data}
\end{center}
\end{figure}
by a Pb-fibre calorimeter of higher granularity,
an MWPC by a planar drift chamber and  adding
a high resolution Silicon strip detector telescope for electron
track identification and kinematic reconstruction. This upgrade permitted
the measurement to be extended
 to high $y \leq 0.89$ in order to access $F_L$
(Section~\ref{FL}) and to low $y \geq 0.003$
 in order to reach the $x$ range
covered by DIS fixed target experiments. Comparing the data
shown in  Fig.~\ref{h1data} with the initial HERA data, Fig.~\ref{92data},
one recognizes the impressive progress made in a few years. The data
are well described by NLO QCD 
 as discussed in Section~\ref{gluon}. Consistent results on
preliminary $F_2$ data were previously
 obtained by the H1 and ZEUS Collaborations~\cite{tony}.

\begin{figure}[htb]
\begin{center}
\begin{picture}(200,220)
\put(-60,-30){
\epsfig{file=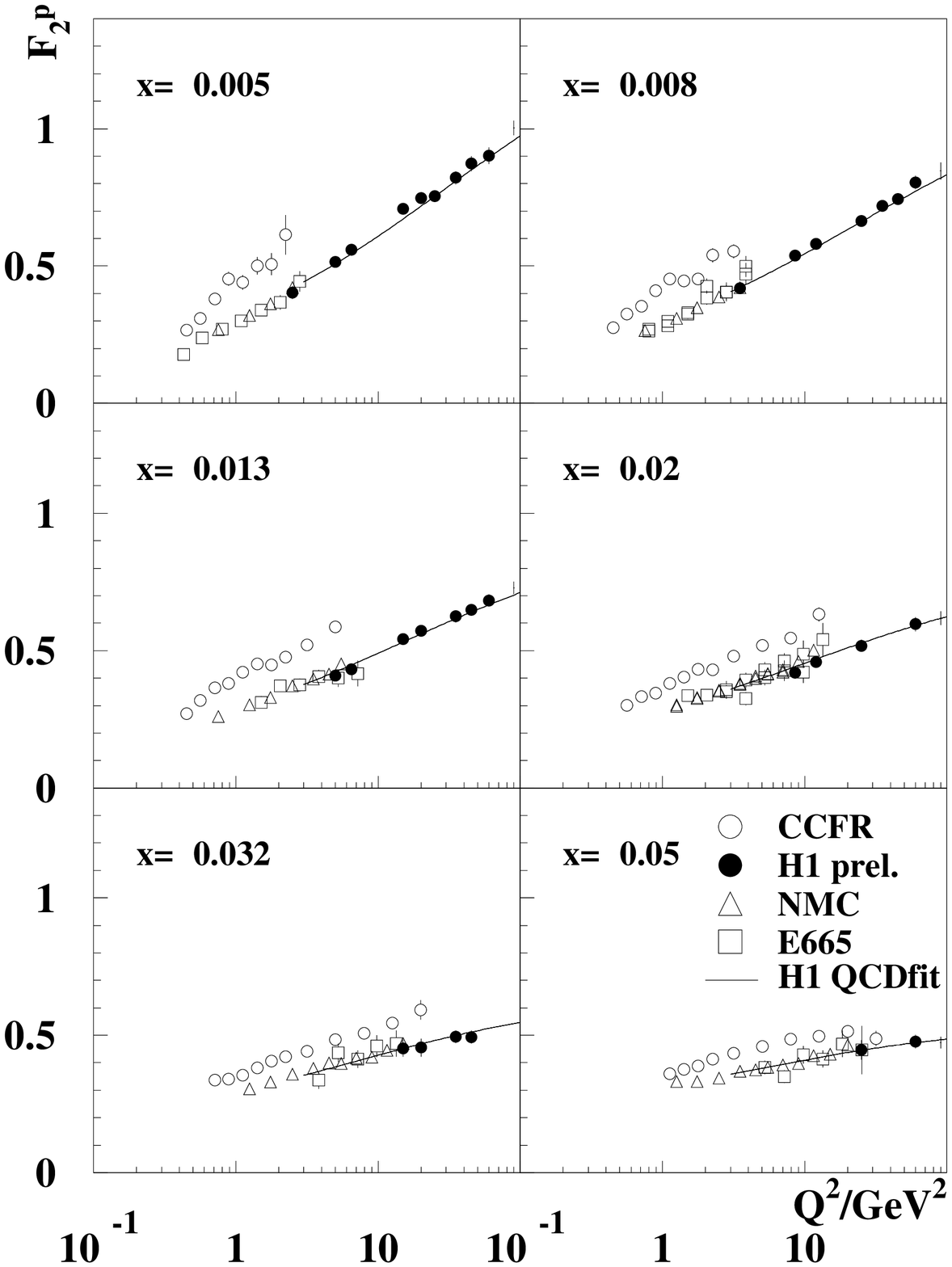,
width=9.5cm,height=10.cm,bbllx=0pt,bblly=0pt,bburx=557pt,bbury=792pt}}
\end{picture}
\caption{$F_2$ structure function data of CCFR, NMC, E665 and H1.
The CCFR data
were corrected for nuclear effects and for the difference of $F_2$ in
charged lepton and neutrino scattering~\cite{yangman}. The
CCFR data are shown with statistical errors only.}
\label{nmccfr}
\end{center}
\end{figure}

The H1 data help resolving a long standing controversy between 
NMC and E665 $\mu p$ data  
and the CCFR $\nu N$  data
on the  structure function $F_2$. As shown in 
Fig.~\ref{nmccfr} the H1 data overlap and extrapolate well to the
$\mu p$ data. The CCFR $F_2$ 
determination which is being redone~\cite{yangman} was recently criticized 
regarding the treatment of charm
and shadowing~\cite{boros}. 
 Since $F_2$ and $xF_3$ add up
to the measured cross section, an $F_2$ reanalysis may  affect also 
the value of $\alpha_s$ derived from $xF_3$. The CCFR cross section
measurement improved in a consistent way
the  CDHSW cross section data. Those seem not to be in contradiction
with muon data~\cite{barone}.
%

Precision measurements at HERA are essential for
calculating 
the expected rates at LHC energies and also permit to estimate
the neutrino scattering cross sections in
active galactic nuclei or gamma ray bursts  at ultra high 
energies, up to $E_{\nu} \simeq 10^{12}$~GeV. Recently 
 very high energy rates were calculated
 using the DGLAP equations~\cite{gandhi}, 
the GRV approach in
DGLAP QCD~\cite{grk} and a combination of DGLAP and BFKL 
dynamics~\cite{kms} which agree remarkably well.
\subsection{Longitudinal Structure Function \boldmath{$F_L$}}
\label{FL}
%
\begin{figure}[htb]
\begin{center}
\begin{picture}(200,200)
\put(-80,-90){
\epsfig{file=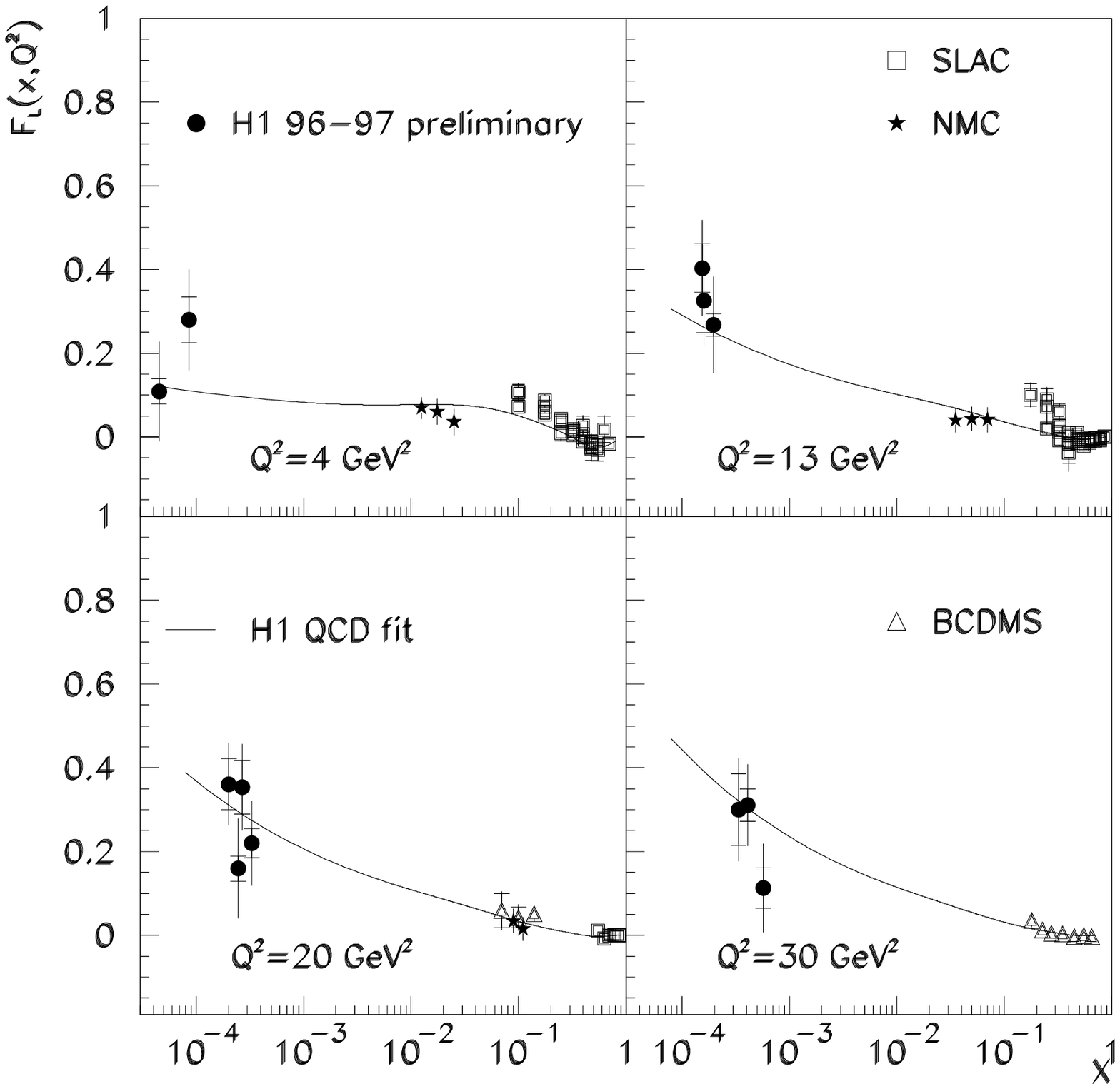,
width=10cm,height=13.5cm,bbllx=0pt,bblly=0pt,bburx=557pt,bbury=792pt}}
\end{picture}
\caption{Measurements of the longitudinal structure function
$F_L(x,Q^2)$ in $ep$ and $\mu p$ scattering. At low $x$ $F_L$
is large because of dominant contributions of
a large gluon momentum density. The four bins comprise data between
$Q^2 = 2, 5, 15, 25$ and 50~GeV$^2$, respectively. The curves represent
the H1 calculation of $F_L$.}
\label{fluta}
\end{center}
\end{figure}
In the naive QPM the longitudinal structure function $F_L$ 
is zero since partons have spin 1/2. In QCD it acquires  
a possibly large value due to gluon emission and represents
together with  $F_2$ a strong constraint to the theory in NLO. 

The sum of $\nu$ and $\bar{\nu}$ nucleon scattering cross sections
is proportional to
        $2xF_1 (1 + \epsilon R) - Y_- \cdot \Delta xF_3 /2Y_+$
and thus is sensitive to $R$
where $\epsilon = 2(1-y)/Y_+$ is the polarization of the 
$W$ boson exchanged and $Y_{\pm}=1 \pm (1-y)^2$.
The CCFR Collaboration has studied the $Q^2$ dependence of $R$
for $0.015 \leq x \leq 0.5$ and $Q^2 < 5$~GeV$^2$
using phenomenological descriptions for
the strange and charm quark distribution
difference determined by $\Delta xF_3 \simeq 4x(s-c)$.
The ratio $R$  tends to be large, $R \geq 0.5$, at small
$Q^2 \simeq 1-2$~GeV$^2$ and $x < 0.1$. For $Q^2 >  10$~GeV$^2$
the function $\Delta xF_3 = xF_3^{\nu} -xF_3^{\bar{\nu}}$
was extracted which is of interest for the
treatment of massive charm~\cite{ccfryang}. 

Using unpolarized targets
the HERMES Collaboration measured the ratio of nitrogen to deuterium
electroproduction cross sections to be astonishingly small at low
$Q^2$ \cite{hermesA}. This effect has been attributed to
a very large ratio $R_N/R_D \geq 5$ in the region
$0.01 < x \leq 0.06$ and $0.5 \leq Q^2 < 1.5$~GeV$^2$
with as yet unexplained origin.

The measurements of the longitudinal structure function in
$ep$ and $\mu p$ scattering are summarized in Fig.~\ref{fluta}.
The H1 data were obtained using assumptions for the behaviour of
$F_2$ in QCD (for $Q^2 > 10$~GeV$^2$) and, independently of QCD,
for the derivative
$\partial F_2 / \partial \ln y$ (for $Q^2 < 10$~GeV$^2$) in the
high $y$ region~\cite{mkvanc} where the cross section approaches $F_2 - F_L$.
Contrary to fixed target experiments such assumptions are
possible since HERA covers more than two orders of
magnitude in $y$ where $F_2$ can be fixed independently of $F_L$. 
The overall behaviour of $F_L$ as a function of $x$ is well described by
a QCD fit in NLO using $F_2$ data only, i.e. by deriving the gluon 
(and parton) distributions from scaling violations and then 
calculating  $F_L$ (Fig.~\ref{fluta}).

The behaviour of $R$ observed at low $Q^2 \simeq 1$~GeV$^2$ 
and the so far limited accuracy of the H1 $F_L$ data, obtained with 
6.8~pb$^{-1}$, represent a challenge for forthcoming
experiments and their theoretical interpretation.
This comprises the  
hypothesis of particularly large higher twist effects  
and large higher order corrections which at low $x$ and $Q^2$
may become even negative in NLO due to
a large negative contribution of the  gluonic 
coefficient function~\cite{willyfl}.
\subsection{Weak Neutral Currents at HERA}
\begin{figure}[htb]
\begin{center}
\begin{picture}(200,200)
\put(-140,-70){
\epsfig{file=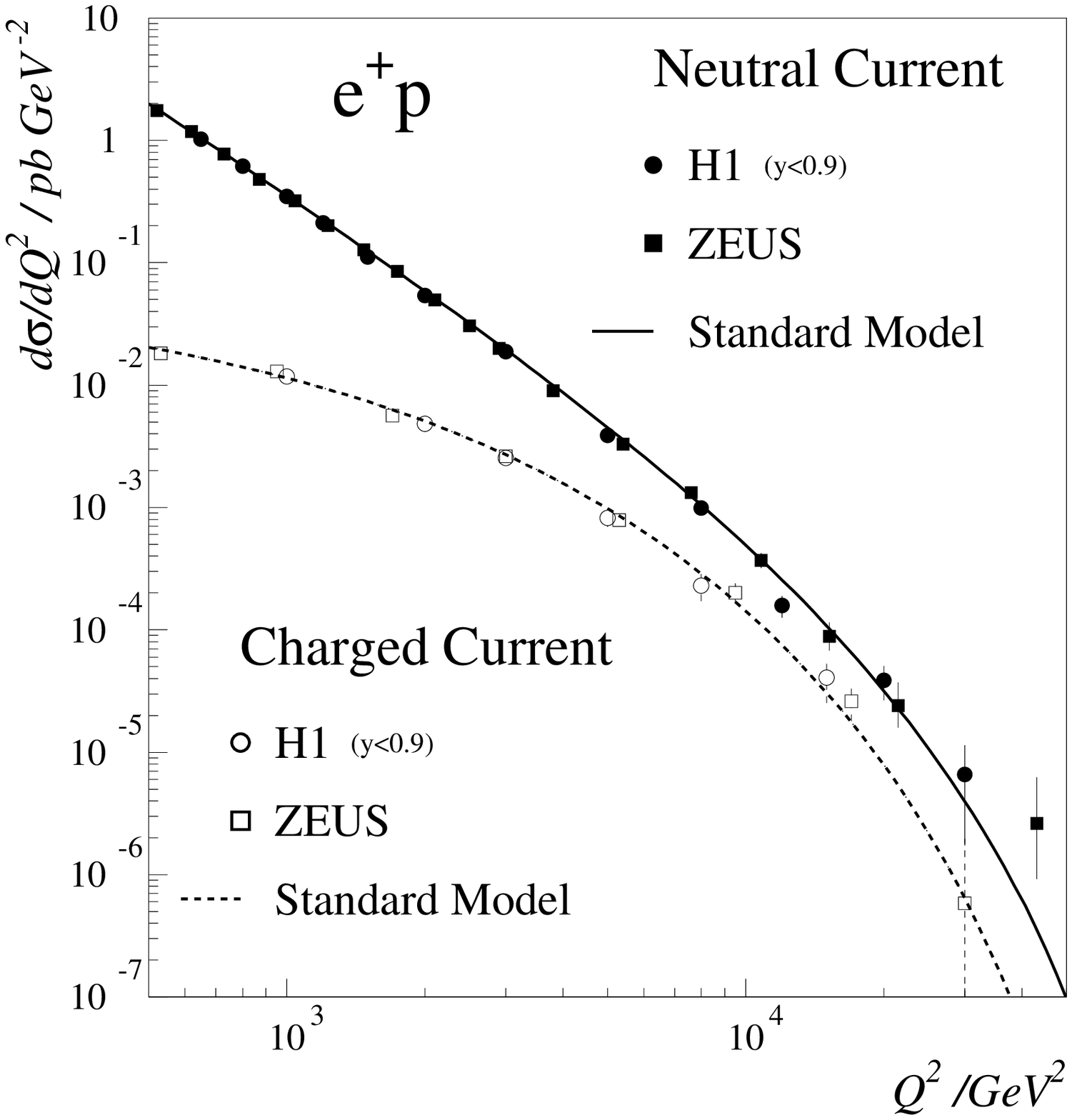,
width=8.5cm,height=12cm,bbllx=0pt,bblly=0pt,bburx=557pt,bbury=792pt}}
\put(90,-70){
\epsfig{file=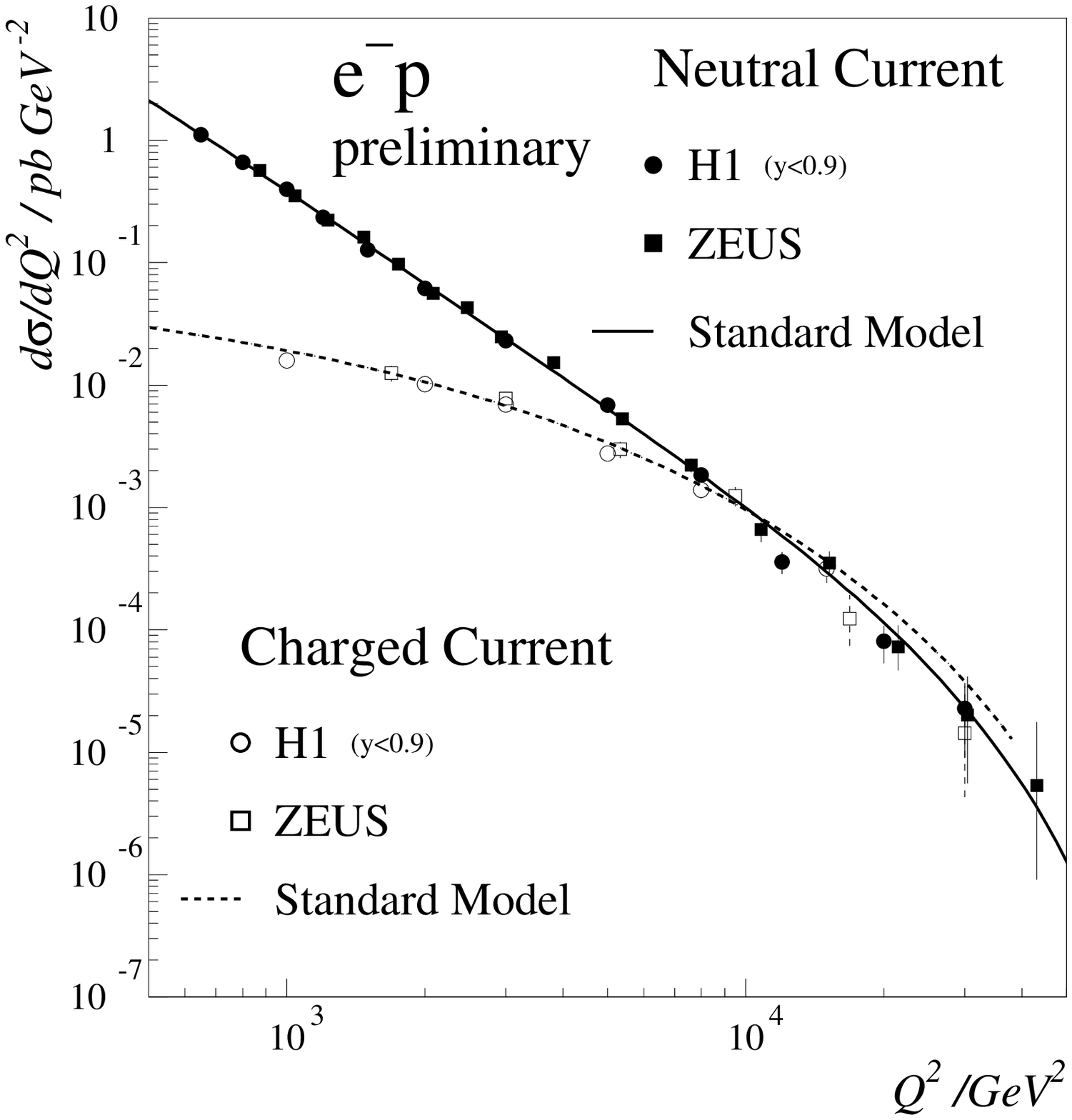,
width=8.2cm,height=12cm,bbllx=0pt,bblly=0pt,bburx=557pt,bbury=792pt}}
\end{picture}
\caption{
Measurements of the $Q^2$ dependence of the positron and the electron
proton neutral and charged current scattering cross sections at HERA,
using data taken in 1994-1997 ($e^+ ,E_p=820$~GeV)
and in 1998-1999 ($e^- ,E_p=920$~GeV). Electromagnetic and weak
interaction cross sections become of similar strength for
$Q^2 \simeq M_Z^2, M_W^2$.}
\label{ncccq}
\end{center}
\end{figure}

At high $Q^2 \simeq M_{W,Z}^2$ photon, $Z$-boson and $W$-boson
exchange are of comparable strength. Thus electroweak interactions
can be used to probe proton structure in neutral (NC) and charged
current (CC) scattering at HERA in the same experiments.
This is demonstrated with the $Q^2$ distributions in electron
and positron proton NC and CC scattering, Fig.~\ref{ncccq},
measured by H1 ($e^+$ NC, CC~\cite{ncplu}; 
$e^-$ NC, CC~\cite{ncmi}) and by ZEUS ($e^+$ NC \cite{zeuseplnc},
$e^+$ CC \cite{zeuseplcc} and  $e^-$ NC, CC \cite{zeusminccc}).

 The double-differential
NC cross section, neglecting the three longitudinal 
structure functions, is given by two generalized structure functions
${\bf F_2}$ and ${\bf xF_3}$
\begin{equation}
        \frac{d^2\sigma^{\pm}}{dQ^2dx} = \sigma^{\pm} = 
        \frac{2 \pi \alpha^2}{Q^4 x} \cdot [Y_+{\bf F_2^{\pm}}
         + Y_-{\bf xF_3^{\pm}}].
        \label{signc}
\end{equation}

These depend on the quark couplings and distributions
but, contrary to hadronic tensor definitions of
structure functions~\cite{derman},
they depend also on the weak electron couplings $v,~a$ to the $Z$ boson,
on the longitudinal electron beam polarization ($\lambda$)
and  on the  propagators via 
$\kappa=Q^2/[4 \sin^2{\theta_W} \cos^2\theta_W (Q^2 + M_Z^2)]$
where $\theta_W$ is the electroweak mixing angle.  They comprise
five genuine structure functions~\cite{mkt}
\begin{eqnarray}
        {\bf F_2^{\pm}}~  = F_2 + \kappa (-v \mp \lambda a) G_2
    + \kappa^2 (v^2 +a^2 \pm 2 \lambda a v) H_2
        \label{bf2} \\
        {\bf xF_3^{\pm}} =   \kappa (\lambda v \pm a) xG_3
    + \kappa^2 (- \lambda(v^2 +a^2) \mp 2 a v) xH_3,
        \label{bxf3}
\end{eqnarray}
defined in Section~1, Eqs.~\ref{fgh} and \ref{gh}.
The ${\bf xF_3}$ term ($\propto Y_-$)
contributes sizeably only at large $y$ and high $Q^2$.  The 
high $Q^2$ NC cross sections measured 
currently at HERA for $\lambda = 0$ are approximately given by
\begin{equation}
      \sigma^{\pm} \simeq Y_+ \cdot F_2 \pm \kappa a Y_- \cdot xG_3. 
      \label{ncapp}
\end{equation}
This causes a positive charge asymmetry between electron and positron
scattering which is proportional to $a a_q$, i.e. parity conserving,
and which is 
determined by the function $xG_3$ measured previously by BCDMS
at lower $Q^2$ for an isoscalar target, see Fig.~\ref{xg3}.

\begin{figure}[htb]
\begin{center}
\begin{picture}(200,320)
\put(-110,-10){
\epsfig{file=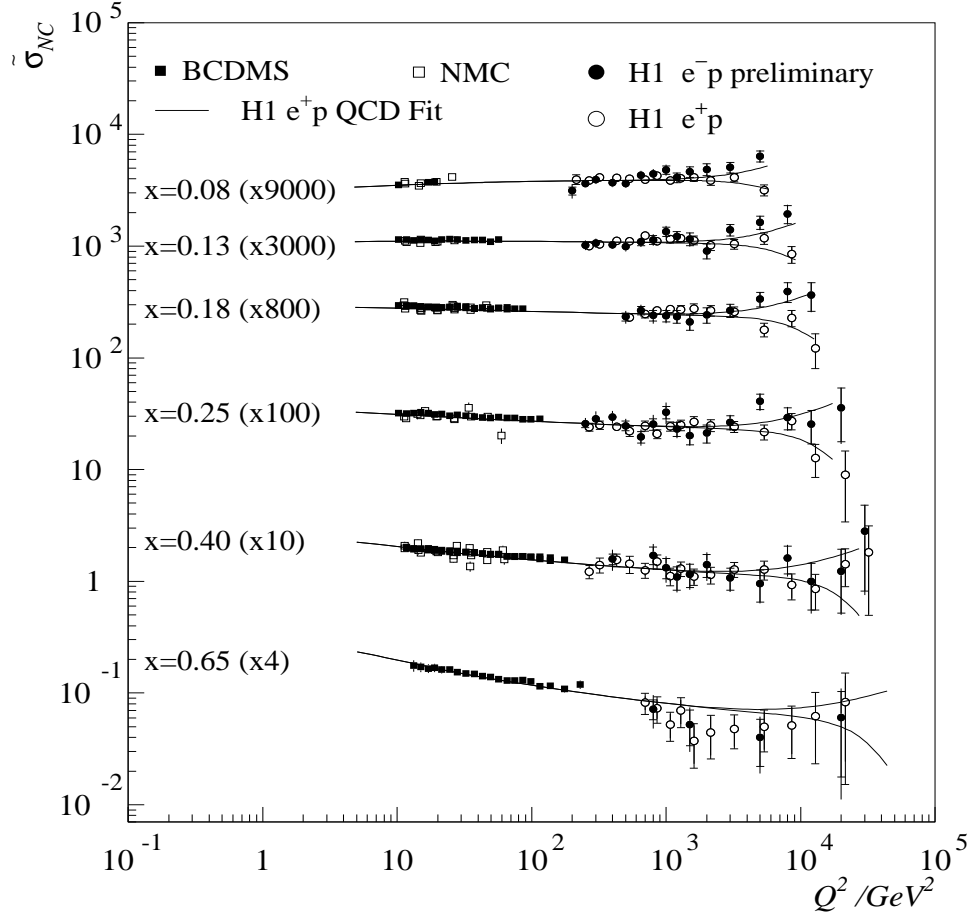,
width=14cm,height=18.5cm,bbllx=0pt,bblly=0pt,bburx=557pt,bbury=792pt}}
\end{picture}
\caption{Measurements of the double differential NC  $e^+$ and $e^-$
proton scattering cross sections by H1 and NMC and BCDMS 
data. The high $Q^2$ H1 data are well described by a QCD fit,
which may  even be restricted
to $Q^2 < 150$~GeV$^2$~\cite{ncmi}, representing a 
remarkable confirmation of the DGLAP evolution in $Q^2$.
A fit to solely H1 and NMC data passes 
through the H1 points
but undershoots the BCDMS data at largest $x$
significantly.
}
\label{h1g3}
\end{center}
\end{figure}

The H1 Collaboration has performed 
measurements of double differential NC scattering cross 
sections~\cite{ncmi} using 35.6~pb$^{-1}$ of $e^+$ data~\cite{ncplu}
taken in  1994-97 at $E_p=820$~GeV and 15~pb$^{-1}$ 
of $e^-$ data~\cite{ncmi} 
taken in  1998-99 at $E_p=920$~GeV.  A comparison
of the cross section measurements with electrons and 
positrons is illustrated in Fig.~\ref{h1g3} which agrees with expectation
based on the $\gamma Z$ interference in NC scattering.
\newpage
\section{Light and Charm Quark Distributions}
\label{lcqd}
\subsection{Charged Currents and Up and Down Quarks}
\label{updo}
New information on the up and down quark distributions became 
available from improved measurements of the charged current cross section
at HERA by H1 and ZEUS. The double-differential CC scattering
cross section is given as 
\begin{equation}
\frac{d^2 \sigma_{cc}^{\pm}}{dx dy} = \frac{G^2}{2 \pi} \cdot
 \left(\frac{M_W^2}{Q^2 + M_W^2}\right)^2 \cdot s \frac{1 \pm \lambda}{2}
  \cdot [Y_+ W_2^{\pm} \mp Y_- xW_3^{\pm}]
        \label{sigcc}
\end{equation}
where $G$ is the Fermi constant and $M_W$ the mass of the $W$ boson.
The CC cross section
 contains two structure functions for a given lepton beam charge and
is proportional to $s$. The HERA energy is equivalent to 53.9~TeV neutrino
beam energy in a neutrino-nucleon fixed target experiment.
 The energy dependence is 
damped for $Q^2 \geq M_W^2$.  In the QPM the CC structure functions 
are combinations of up and down quark distribution sums, i.e.
$W_2^+=D+\bar{U}$, $W_2^-=U+\bar{D}$, $xW_3^+=D-\bar{U}$      
and $xW_3^-=U-\bar{D}$ with $U = x(u+c)$ and $D=x(d+s)$. 
At large $x \geq 0.3$
the valence quark distributions $u_v$ and $d_v$ dominate the
interaction cross sections, i.e.
\begin{eqnarray}
        \sigma(e^+p \rightarrow \bar{\nu}X)~~\propto~~
         \bar{U} + (1-y)^2 D \rightarrow (1-u)^2 xd_v
\\
         \sigma(e^-p \rightarrow \nu X)~~\propto~~
         U + (1-y)^2 \bar{D} \rightarrow ~~~~~~~~~~xu_v
        \label{sclim} 
\end{eqnarray}
for $x \rightarrow 1$. A complete set of
double differential $e^{\pm}p$ CC cross section data
was presented by H1 using 
36~pb$^{-1}$ of positron-proton data (1994-1997)~\cite{ncplu}
and 15~pb$^{-1}$
of electron data (1998-1999)~\cite{ncmi}.
\begin{figure}[htb]
\begin{center}
\begin{picture}(200,260)
\put(-80,-15){
\epsfig{file=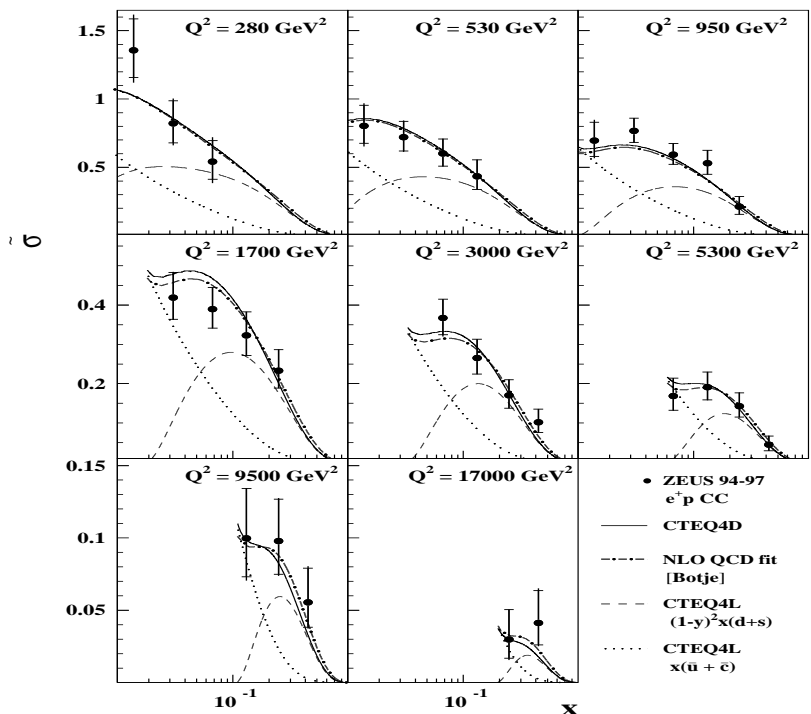,
width=10cm,height=9.5cm,bbllx=0pt,bblly=0pt,bburx=557pt,bbury=792pt}}
\end{picture}
\caption{
ZEUS measurement of the CC  $e^+p$ scattering
cross section compared with $\bar{U}$ and $D$
 quark distributions, see Eq.~\ref{sclim}, and different QCD fits.}
\label{cch1}
\end{center}
\end{figure}
%
The $U$ dominated $e^-p$
cross section was found to be about 5 times larger than the $e^+p$
cross section at $Q^2 \simeq 10,000$~GeV$^2$.
The  $e^+p$ CC data of H1 are consistent
with the published measurement of the ZEUS
Collaboration~\cite{zeuseplcc} based
on 47.7~pb$^{-1}$, Fig.~\ref{cch1}. 
The NC and CC measurements at high $Q^2$ are of particular interest 
for the determination of the $d/u$ ratio at high $x$
because their interpretation is free of nuclear corrections.
Yet, an order of magnitude increase in luminosity is still
required to access the high $x$ region
which represents one of the goals of the HERA luminosity upgrade
programme.

Deuterium binding corrections were recently reconsidered,
and $d_v$ was adjusted to be larger than previously assumed~\cite{bounk},
the ratio $d_v/u_v$ for $x \rightarrow 1$ tending to 0.2.
An enlarged $d$ quark distribution fits to the 
$W^{\pm}$ charge asymmetry data in $p \bar{p}$ collisons.
Violation of $u$ and $d$ quark symmetry in protons and neutrons,
however, which was suggested to explain
the difference between the CCFR and NMC $F_2$ data~\cite{borcsv},
leads to too large a $W$ asymmetry~\cite{bodekdis}.   
\subsection{Sea Quarks}
\begin{figure}[htb]
\begin{center}
\begin{picture}(200,175)
\put(-120,-10){
\epsfig{file=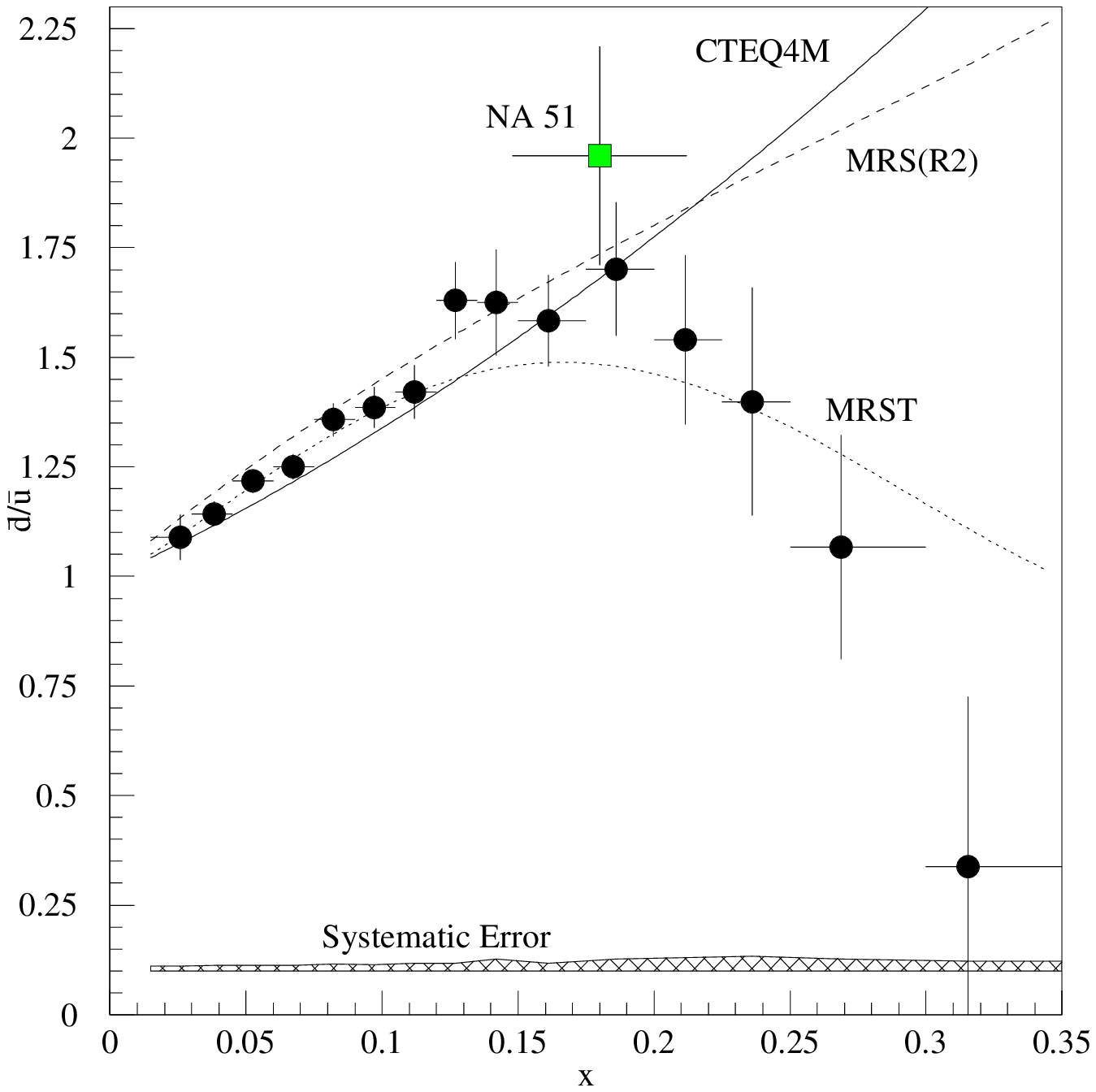,
width=9.5cm,height=13.5cm,bbllx=0pt,bblly=0pt,bburx=557pt,bbury=792pt}}
\put(100,-13){
\epsfig{file=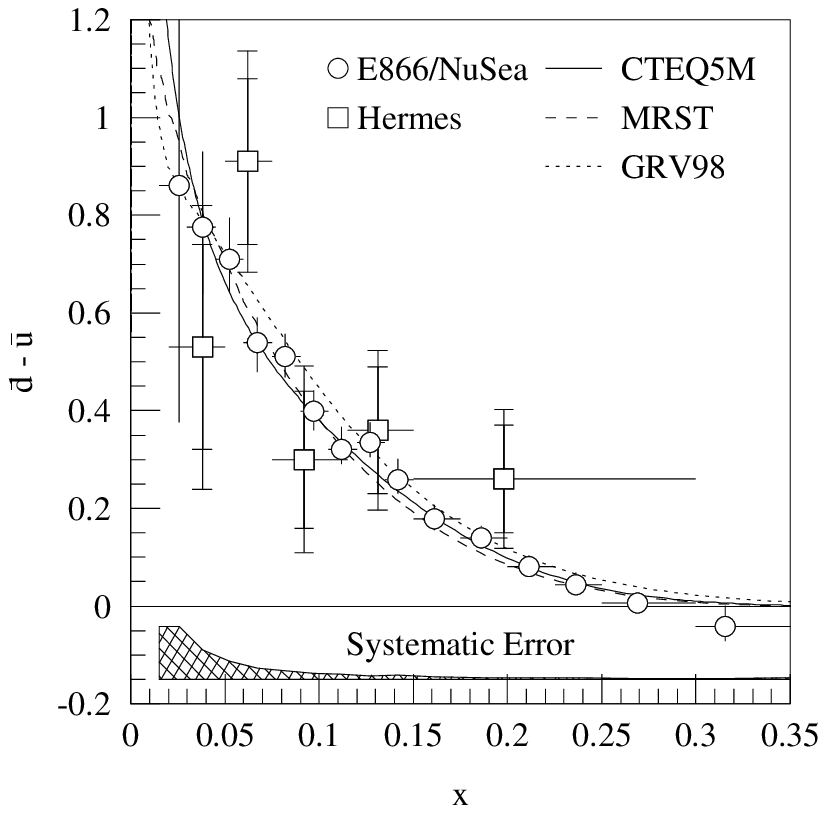,
width=17cm,height=23cm,bbllx=0pt,bblly=0pt,bburx=557pt,bbury=792pt}}
\end{picture}
\caption{Measurements of $\bar{d}/\bar{u}$  
and of $\bar{d}$ -$\bar{u}$ by NA 51, E866/NuSea and by HERMES
compared with recent structure function fits.}
\label{dbaroubar}
\end{center}
\end{figure}

Interesting data become available on the flavour asymmetry
in the nucleon sea. From a high statistics measurement of
Drell-Yan muon pair production in $pp$ and $pd$ collisions at the
Tevatron, the E866/NuSea Collaboration obtained
for 
$\int_{0}^{1} (\bar{u} -\bar{d}) dx$
 a value of
-0.118 $\pm$ 0.011 
at  $ \langle Q^2 \rangle = 54$~GeV~$^2$~\cite{handtuch}.
This confirms and also significantly
improves the previous NMC result of $-0.15 \pm 0.04$ which was
derived from a measurement of the Gottfried sum rule
$\int_{0}^{1} [(F_2^p - F_2^n)/x] dx = 
1/3 + 2/3 \cdot \int_{0}^{1} (\bar{u} -\bar{d}) dx $.
The measured ratio $\bar{d}/\bar{u}$ as a function of 
$x$ is shown in Fig.~\ref{dbaroubar}. The data have considerable 
impact on global parametrizations of parton distributions.  
%
 A consistent result, albeit of less statistical accuracy, was
 obtained by the HERMES Collaboration~\cite{hermespi}
  with a measurement of
 semi-inclusive $\pi^{\pm}$ production in unpolarized
 $ep$ and $ed$ scattering at lower $\langle Q^2 \rangle = 2.3$~GeV$^2$. 
 A violation of flavour symmetry is not predicted in perturbative
 QCD which points to non-perturbative effects such as Pauli blocking
 and pion clouds. In the latter model the nucleon is expanded in a 
 Fock state of mesons and baryons. Phenomenologically
 one finds more $\pi^+$ than $\pi^-$ in the nucleon with a momentum
 distribution peaking at $x_{\pi} \simeq 0.2$~\cite{scur}.
 
The NuTeV Collaboration~\cite{nutev} determined the strange quark
distribution to be about 1/2 of the averaged nucleon sea, i.e.
$s = [0.42 \pm 0.07 (syst) \pm 0.06 (stat)] \cdot 
(\bar{u}-\bar{d})/2$, in agreement with previous 
analyses of dimuon production in neutrino-nucleon scattering
experiments.

Indications for a difference of the strange and anti-strange quark 
distributions  at large $x \simeq 0.6$ 
were obtained in a recent reanalysis and global fit
of DIS and Drell-Yan data~\cite{barone}. Sensitivity to
$(s-\bar{s})$ in this analysis comes from the CDHS data
measuring $\sigma^{\nu} - \sigma^{\bar{\nu}} \propto
x(s-\bar{s}) + Y_- x(u_v+d_v)$ at high $x$.
Such a strange asymmetry
is possible in models considering states as $K^+ \Lambda$ to be intrinsic
to the nucleon where $K^+$ yields $\bar{s} \propto (1-x)$
and $\Lambda$ yields $s \propto (1-x)^3$~\cite{melni}.
\subsection{Charm}
\label{charm}
%
\begin{figure}[htb]
\begin{center}
\begin{picture}(200,240)
\put(-130,-10){
\epsfig{file=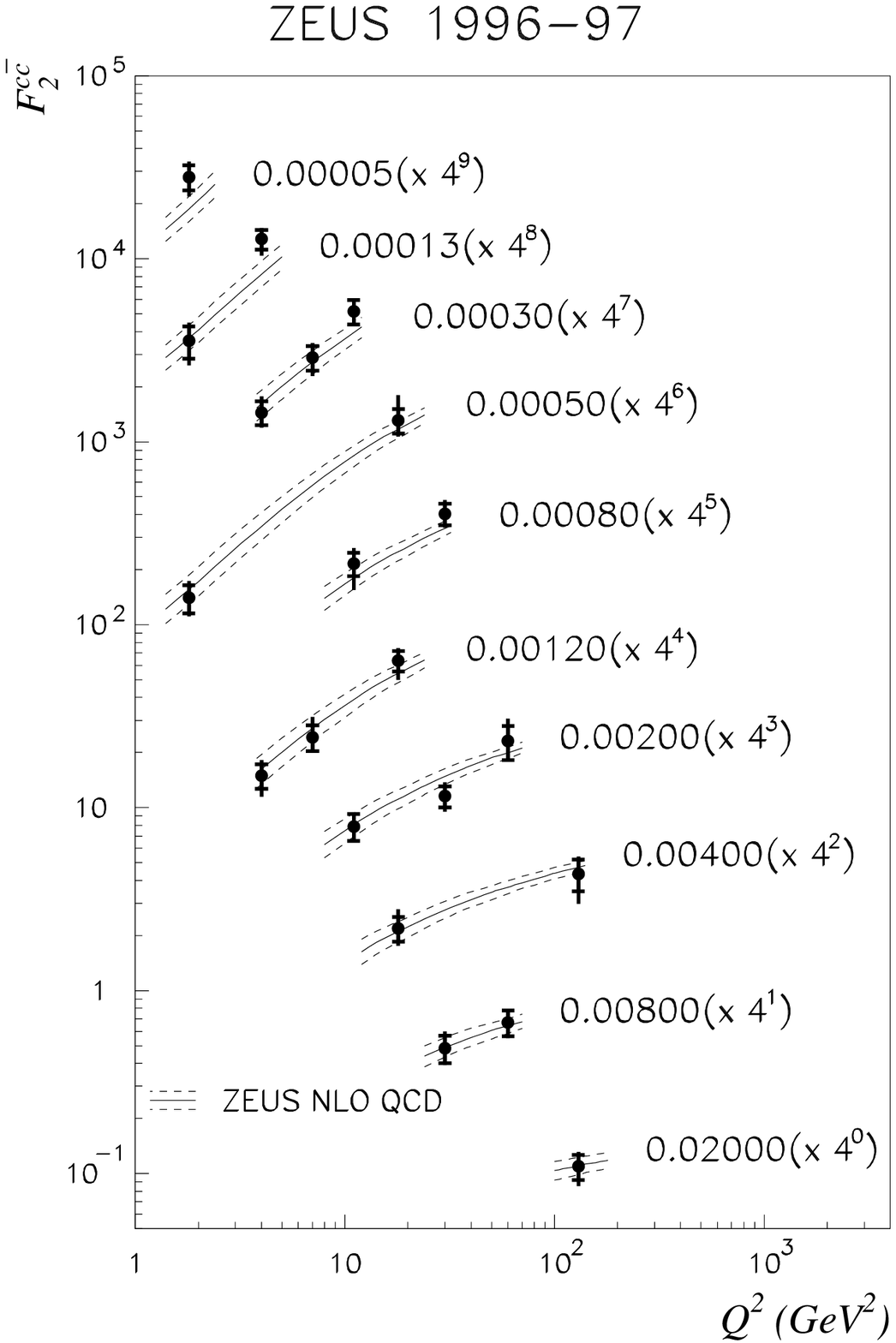,
width=10.5cm,height=12.5cm,bbllx=0pt,bblly=0pt,bburx=557pt,bbury=792pt}}
\put(85,-10){
\epsfig{file=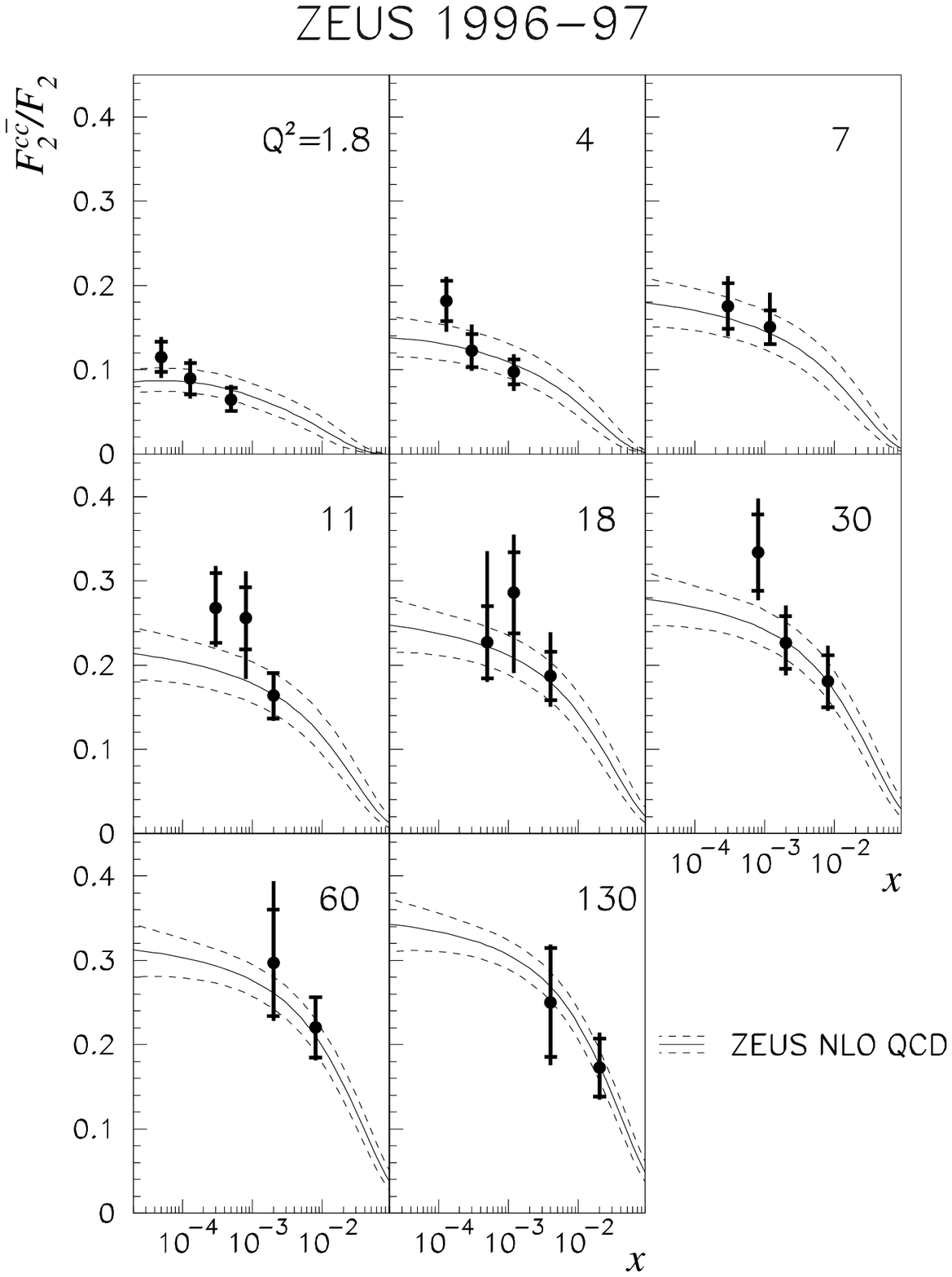,
width=10.5cm,height=12.5cm,bbllx=0pt,bblly=0pt,bburx=557pt,bbury=792pt}}
\end{picture}
\caption{Measurement of $F_2^{c\bar{c}}(x,Q^2)$ and of
the ratio $F_2^{c\bar{c}}/F_2$ in $ep$ scattering at HERA
by the ZEUS Collaboration with 37~pb$^{-1}$ of data. The dashed error
bands denote the uncertainty of the QCD fit which is dominated by the
charm quark mass range chosen to be 1.2 to 1.6~GeV.}
\label{zeusfcc}
\end{center}
\end{figure}
Charm, as was already noticed by Witten  in 1976, may ``subject
non-Abelian theories to a rigorous experimental test by measuring 
the charmed quark contribution to structure 
functions''~\cite{extrinsic}.
Since then the charm and beauty treatment in perturbative QCD
 has been worked out to 
higher orders~\cite{smith}.
Variable flavour schemes are being studied~\cite{tungvf}
to correctly handle the heavy flavour contributions
near and beyond threshold in 
 analyses of parton distributions,
of the gluon distribution and of $\alpha_s$.
A new measurement
of the charm structure function $F_2^{c \bar{c}}$ was
performed by the ZEUS Collaboration~\cite{fcczeus}  
 using the $\Delta M$ tagging 
technique for $D^* \rightarrow K2\pi$ and $K4\pi$, Fig.~\ref{zeusfcc}.
The relative contribution of charm is large, reaching
30\% at low $x < 0.001$ for  $Q^2 \simeq 100$~GeV$^2$.
This large fraction is due to photon-gluon fusion as
the dominant process for charm production. 
Further experimental progress at HERA towards high precision
will be achieved with new or
upgraded Silicon vertex detectors, higher luminosity, inclusion
of further final states and dedicated track triggers.
\section{Gluon Distribution and Coupling Constant \boldmath{$\alpha_s$}}
\subsection{Scaling Violations at Low  \boldmath{$x$}}
\label{qcd}
Scaling violations in the DIS $Q^2$ region down to low $x \simeq 0.00005$
can be successfully described in the DGLAP formalism.
This is again demonstrated 
with the new precise cross section measurement of H1, 
 Fig.~\ref{h1data}. Conventional QCD fits 
 use  parametrizations of parton distributions at a starting 
 scale $Q_o^2$ and evolve them in $Q^2$ to highest $Q^2 \geq
 M_Z^2$ values up to order $\alpha_s^2$. However,
 the splitting functions
 have expansions which contain also powers of $\ln (1/x)$.
 These are large at low $x$, such that $\alpha_s \ln(1/x) \simeq 1$,
and yet do not  seem necessary to 
 phenomenologically describe the observed structure function behaviour.
 Calculations are performed in order to account for these
$\ln (1/x)$ terms~\cite{alta}  and 
 to cure perhaps the instability of the BFKL equation in NLO~\cite{ciaf}.
 Indications were reported for the presence of
$\ln(1/x)$ terms in inclusive DIS data~\cite{thorne}.
  Experimentally even higher precision is
both  required and possible for the 
 structure function measurements, including
 $F_L$, which may lead to crucial tests of QCD at low $x$. 
Due to unitarity constraints one expects to find saturation of
the rising behaviour of $F_2$ which, however, seems to be beyond the 
low $x$ range accessible by HERA in the DIS region.

\begin{figure}[htb]
\begin{center}
\begin{picture}(200,230)
\put(-130,-160){
\epsfig{file=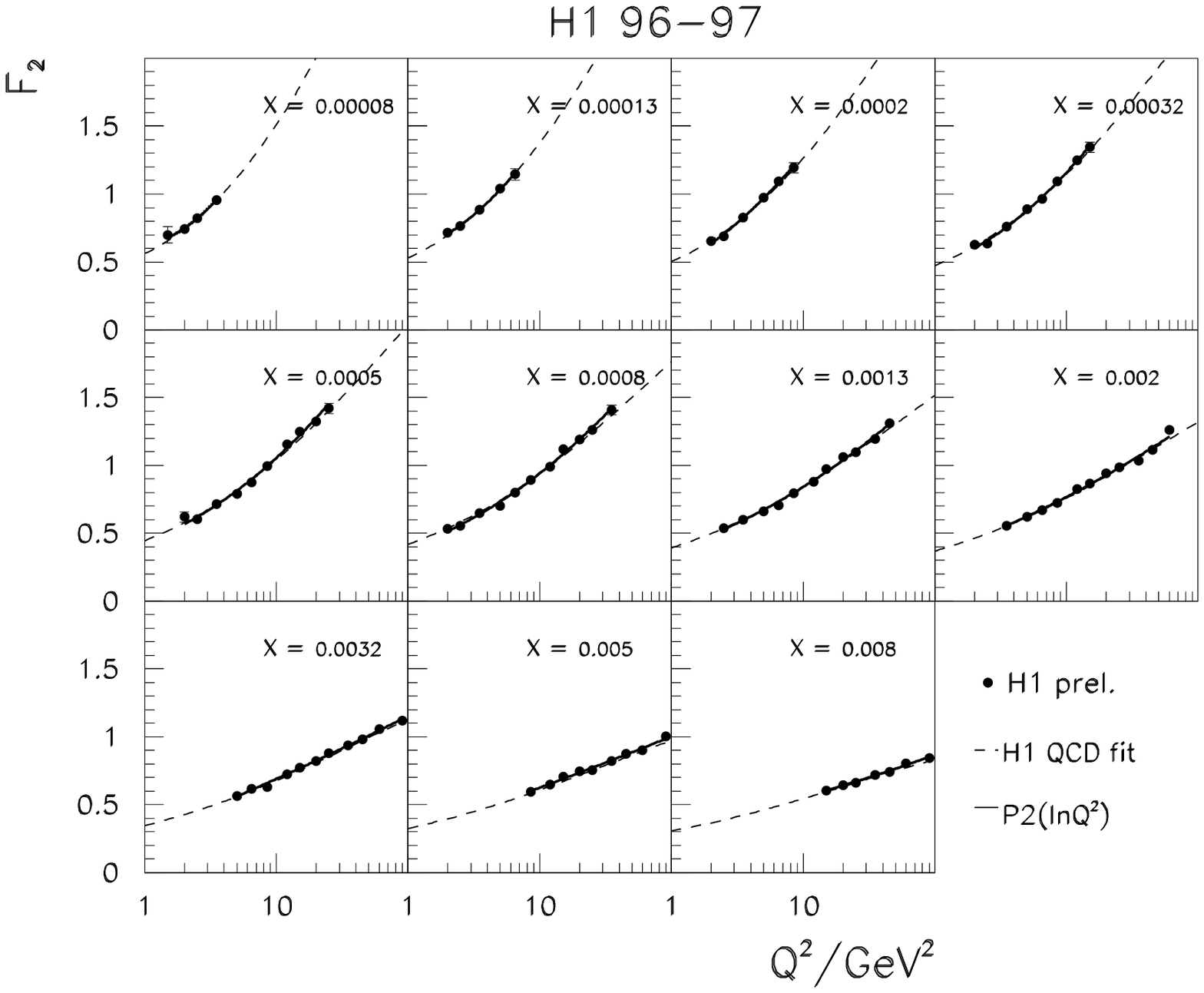,
width=9cm,height=16.5cm,bbllx=0pt,bblly=0pt,bburx=557pt,bbury=792pt}}
\put(110,-70){
\epsfig{file=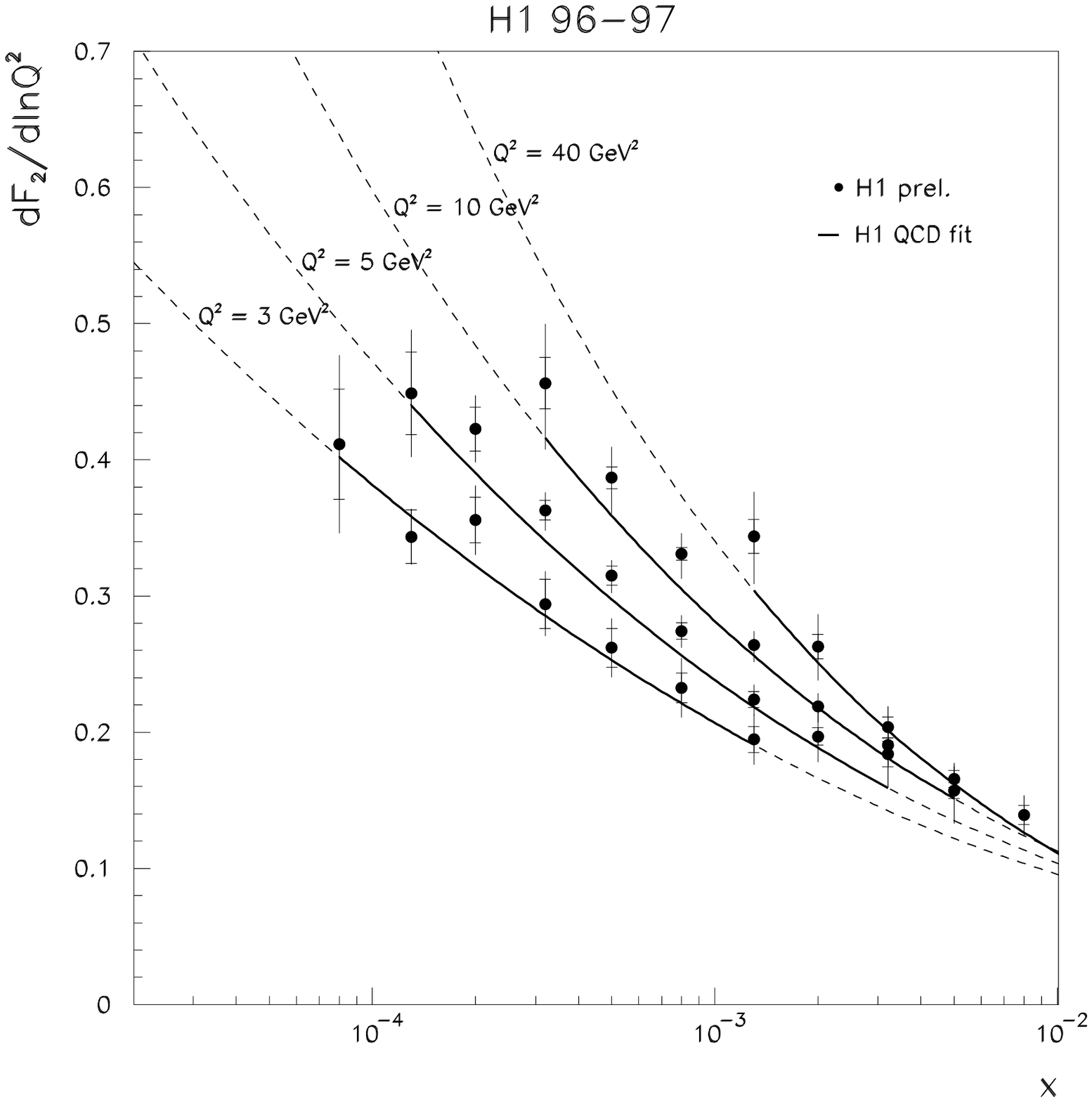,
width=7cm,height=12.5cm,bbllx=0pt,bblly=0pt,bburx=557pt,bbury=792pt}}
\end{picture}
\caption{New preliminary H1 data show  $F_2(x,Q^2)$ to be non-linear
in $\ln Q^2$ at low $x$ (left). The
derivative $\partial F_2 / \partial \ln Q^2$
is a continuously falling function of $x$
for $Q^2 \geq 3$~GeV$^2$ (right). }
\label{f2der}
\end{center}
\end{figure}

Scaling violations are conveniently studied using the $\ln Q^2$
derivative of $F_2$. In Fig.~\ref{f2der} the structure function 
$F_2$ from H1 is shown as a function 
of $Q^2$ for $x < 0.01$. The $\ln Q^2$ dependence is non-linear and 
can be  well described by a quadratic expression
 $P_2 = a + b \ln Q^2 + c (\ln Q^2)^2$ (solid lines) which nearly
coincides with the NLO QCD fit (dashed lines).
The  local derivatives 
$\partial F_2 / \partial \ln Q^2$ determined from the new  H1
$F_2$ data are not constant in $Q^2$ and also depend
on $x$. Approximately they can be described for each bin of $x$ 
by $b + 2 \cdot c \ln Q^2$. Small
deviations from this behaviour occur in NLO QCD.
Using this expression the derivatives are determined
at fixed $Q^2$ and displayed as functions of $x$ 
in Fig.~\ref{f2der}. There is no departure observed from a rising behaviour
of the $\ln Q^2$ derivatives down to $Q^2 = 3$~GeV$^2$. If such a plot 
is made as a one-dimensional distribution,
using the derivatives calculated for each
bin of $x$ at the mean $Q^2$ of a given bin, then the derivative 
$d F_2 / d \ln Q^2$ flattens starting at $Q^2 \simeq 6$~GeV$^2$~\cite{alan}. 
In the region covered by the H1 data this behaviour reflects 
the restriction of the kinematic range of the measurement.  Some analyses
of the ZEUS data extending to lower $Q^2 \simeq 1$~GeV$^2$
introduce screening corrections 
in order to describe the behaviour of 
$F_2$~\cite{gots,mkp}. Both $F_2$ and $F_L$ in this region should be
measured with still higher accuracy (see Section 2.1.)
as these permit important information to be deduced on the 
 dynamic interplay  of gluon and sea distributions, 
on the effect of higher order and power corrections and on the 
shadowing phenomenon.
 
%
\subsection{Gluon Distributions}
\label{gluon}
\begin{figure}[htb]
\begin{center}
\begin{picture}(200,200)
\put(-125,-10){
\epsfig{file=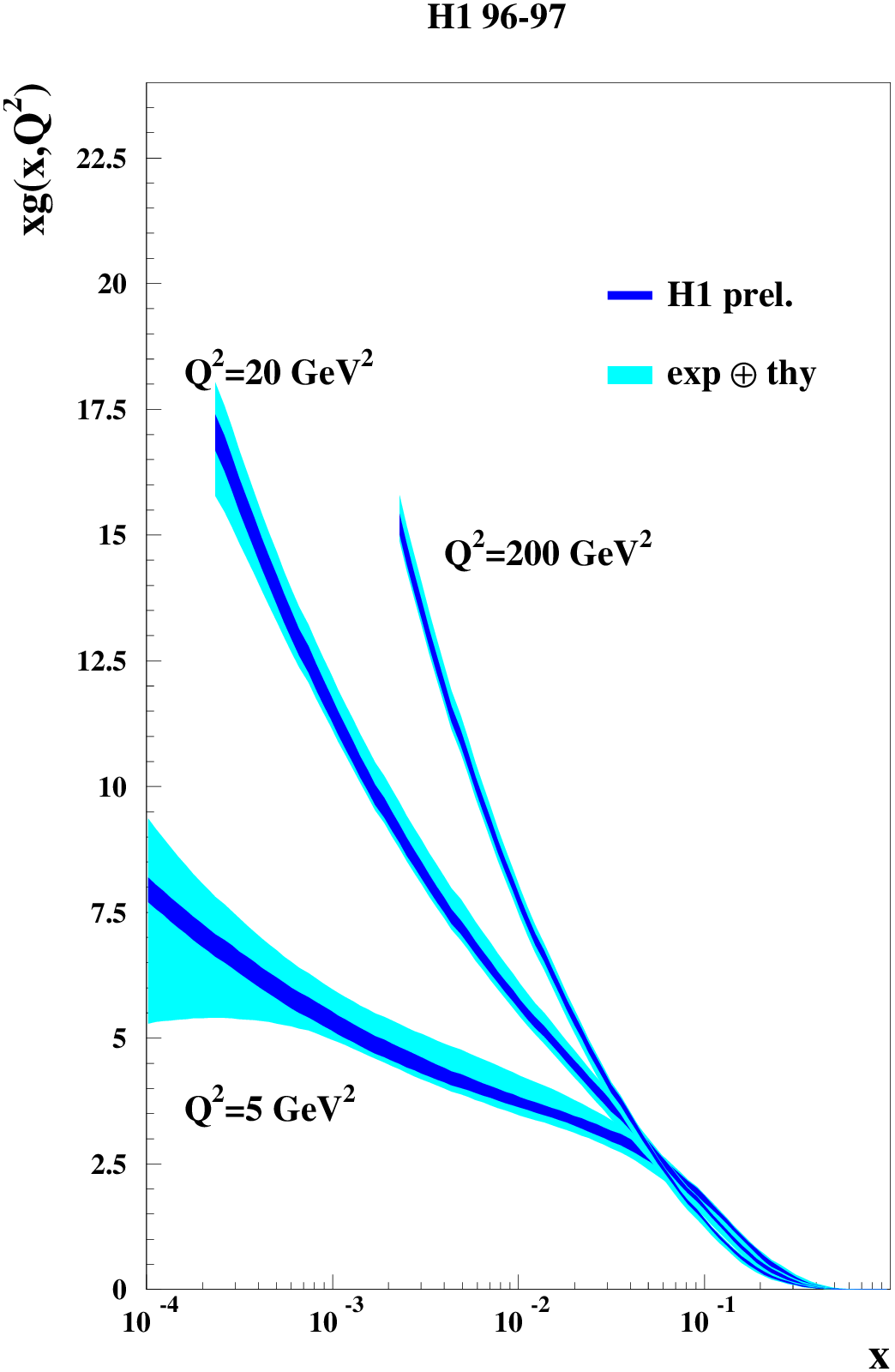,
width=8cm,height=7.3cm,bbllx=0pt,bblly=0pt,bburx=557pt,bbury=792pt}}
\put(90,-50){
\epsfig{file=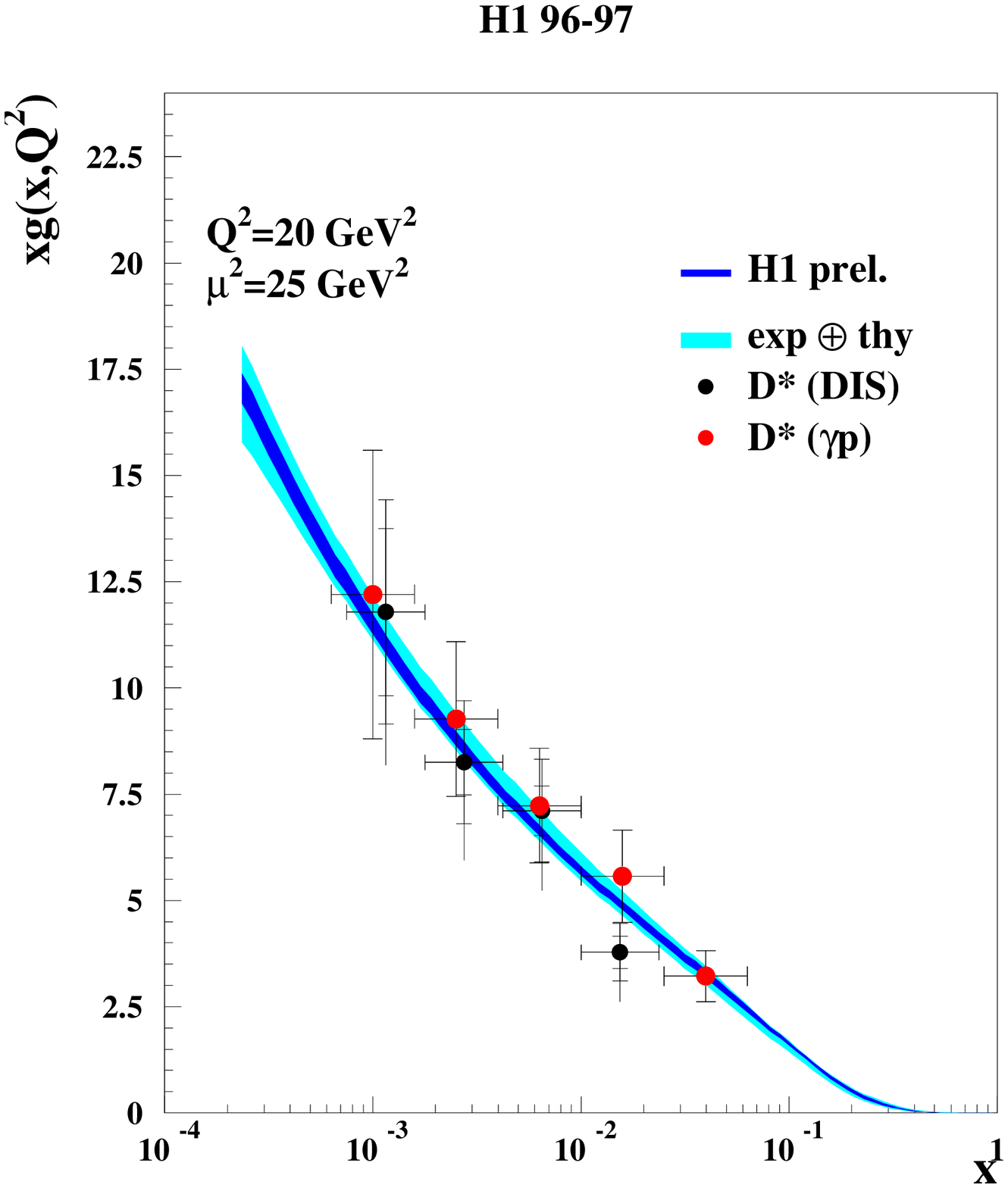,
width=8cm,height=9.9cm,bbllx=0pt,bblly=0pt,bburx=557pt,bbury=792pt}}
\end{picture}
\caption{Determination of $xg$ by H1 using NMC and H1
$lp$ data in NLO QCD  (left). Comparison of $xg$ from scaling violations
with the unfolded gluon distribution from charm $D^*$ measurements 
by H1(right).}
\label{xgh1}
\end{center}
\end{figure}

In QCD the $Q^2$ evolution of $F_2$ is governed by  
 the strong interaction coupling constant $\alpha_s$. 
The evolution relates
 the quark distributions to the gluon distribution
$xg$. The H1 Collaboration has performed
a new NLO QCD fit to the H1 and NMC inclusive cross-section data.
It uses the
DGLAP evolution equations for three light flavours with the charm
and beauty contributions added according to the NLO calculation
of the boson-gluon fusion process \cite{lae2}. 
The proton structure function $F_2$ is
a superposition of two independent functions with different 
evolutions, i.e. $F_2 = 5/18 \cdot S + 1/6 \cdot N$, where 
the singlet function $S=U+D$ is the sum of up and down quark
distributions and the non-singlet function $N=U-D$ is their difference.
In the new H1 fit a different linear combination is introduced
such that $U= 2/3 \cdot V + A$ and $D = 1/3 \cdot V + A$. 
In a simplified parton
model ansatz with $\bu = \bd$ and $s + \bs = (\bu + \bd)/2$
one finds $V = 3/4  \cdot (3 u_v - 2 d_v)$. This allows the
quark counting rule to be applied which constrains $\int{V}dx=3$.
This ansatz is used to 
fit the cross-section data,~Fig.~\ref{h1data},
for $3.5 \leq Q^2 \leq 3000$~GeV$^2$ assuming \amz~$=0.118$.
It is written in the $\overline{MS}$ renormalization scheme and 
generalized to
account for the measured difference  $\bu - \bd$ and the
fraction of strange quarks, see Section~\ref{lcqd}.
The salient feature of this new analysis is that it applies to
DIS proton data only but correctly determines
 the  gluon momentum fraction to 
be about 0.45 at $Q^2 = 10$~GeV$^2$. 
The gluon distribution resulting from  this 
fit is shown in Fig.~\ref{xgh1} (left).
The inner error band defines the experimental uncertainty of a few
per cent at low $x$ using the treatment of 
correlated systematic errors of~\cite{zp}.
The outer error band comprises uncertainties due to dependencies
on the fit parameters ($Q^2_{min}, Q_o^2, \alpha_s, m_c$) and on
the choice of parametrizations for the initial distributions.
A remarkable feature
of $xg$ is the crossing point at $x \simeq 0.06$ which is
analogous to the Bjorken scaling behaviour of $F_2$ and reflects
the conservation of the gluon and quark momenta.
 In Fig.~\ref{xgh1} (right)
the gluon distribution is seen to agree very well with $xg$ unfolded from
the charm structure function DIS and photoproduction
data of H1~\cite{f2ch1} which confirms hard scattering factorization.
 It has early been recognized that 
in photoproduction ($Q^2 \simeq 0$) the charm mass provides
 a hard scale~\cite{misha}.

While $xg$ at low $x$ is well determined by the HERA structure function
measurements, there are sizeable uncertainties of one order of magnitude
at high $x \simeq 0.6$~\cite{begel}.
The gluon distribution is accessed at high 
$x$ by quark-gluon Compton scattering leading to direct photon 
emission~\cite{halzen}.
In a recent experiment by the E706 Collaboration~\cite{e706} the photon 
$p_T$ spectrum was found to exceed QCD expectation by a factor of about two
which has been phenomenologically cured by a Gaussian transverse
momentum smearing with $k_T$ of 1~GeV, larger than the intrinsic 
 $k_T$ value of about 0.4~GeV~\cite{begel}. 
High $E_T$ jet data at large rapidities are sensitive also to $xg$
at large $x$
and lead to a rather high gluon distribution.
Inclusion of different data sets yields remarkably differing results. 
Resolving the issue of $xg$ at high $x$ is essential
for a reliable prediction of  Higgs production 
in $pp$ colliders. It is 
necessary since the high $x$
exponent $c_g$ of $xg \propto (1-x)^{c_g}$ is known to be correlated with
$\alpha_s$.  In this respect
precision measurements of structure functions at high $x$
are important. Since $F_2$ vanishes
as $(1-x)^3$, any measurement error  at large $x$ 
is amplified like $1/(1-x)$. The HERA collider experiments with their
unique possibility to overconstrain the kinematics 
can be expected to lead to precision data also at high $x$~\cite{bkp}
when the luminosity is upgraded. 
 
 Recently  updates  of the GRV 
parametrizations were presented~\cite{grv98}.
 New sets of fits were made by 
the MRST~\cite{mrst99} and the CTEQ 
groups~\cite{cteq5}. GRV98 uses DIS,
$n/p$ and Drell-Yan data assuming \amz=0.114. MRST99 uses  direct
photon data of the WA70 experiment for different $k_T$ and varies 
the $d/u$ ratio, \as~and $m_c$. CTEQ5 does not use direct photon
data but analyzes high $E_T$ jet data instead. Sets are provided for
different renormalization schemes and heavy 
quark treatments. As a consequence there exists a variety  of 
parametrizations illustrating the still large flexibility
of theoretical assumptions and pointing to possible experimental
contradictions. An interesting attempt was made recently~\cite{botje}
to quantify the experimental uncertainties of parton distributions
resulting from global QCD fits to DIS data.
\subsection{Determinations of {\boldmath $\alpha_s$}}
\label{alphas}
New determinations of \amz~with structure function data were
presented recently. Conventional analyses parametrize a set
of input quark distributions and $xg$ at certain input scale 
$Q_o^2$ using the DGLAP equations to NLO to calculate
the theoretical expectation.
Minimization of a $\chi^2$ function determines \as~and 
 the roughly 10-15 parton distribution 
parameters. The treatment of systematic errors
affects both  the central value and the error
size of \amz. At low $Q^2$ power corrections to the logarithmic
evolution may be sizeable and anticorrelate with \as. Since
analyses differ in these assumptions and use different sets of
data, one may not be surprised to still find some spread of
the quoted values of \amz. 
Using   the SLAC, BCDMS and NMC $p$ and $n$
structure function data
and taking into account systematic error correlations
and  higher twists $\propto 1/Q^2$, a value of  
\amz  = 0.1183~$\pm 0.0021 (exp) \pm 0.0013 (thy)$ 
has been derived~\cite{aleck}. A similar analysis~\cite{alfvogt}
including the published HERA data
and adding all errors in quadrature yields
\amz  = 0.114~$\pm 0.002 (exp)^{+0.006}_{-0.004} (thy)$
which is closer to
a previous determination of \amz~based on SLAC and BCDMS 
data~\cite{vm}. The quoted theoretical errors 
represent the uncertainties of the renormalization scale $\mu_r$,
the former analysis compensating part of the $\mu_r$ dependence
with the higher twist contribution. 

The theoretical uncertainties
are diminished in NNLO calculations. 
So far only partial results are available
on the 3-loop splitting functions while the $\beta$ function and
the coefficient functions are known~\cite{willydis}.  
This gave rise to a revival of
moment analyses. In~\cite{kata}
the $xF_3$ data of the CCFR Collaboration
are reconstructed using orthogonal Jacobi polynomials. Power 
corrections are considered and a value of 
\amz = 0.118~$\pm~0.002~(stat)~\pm 0.005~(syst)~\pm 0.003~(thy)$
is obtained in NNLO  corresponding to 0.120 in NLO. While this
uses a pure non-singlet function, not coupled to the gluon distribution,
a new analysis of
SLAC, BCDMS, NMC, ZEUS and H1 data
using  Bernstein polynomials of $F_2$  yields 
\amz = 0.1163~$\pm 0.0023$ in NNLO with a single
 error supposed to comprise all
experimental and theoretical uncertainties. This analysis~\cite{yndu}
is extended to a $Q^2$ range of 2.5 to 230~GeV$^2$ and includes
power corrections. Its NLO
result is 0.1175, and moments of $xg$ are determined. 
 
Although all these analyses represent quite remarkable theoretical and
experimental progress, one still has to be cautious. 
The systematic
error treatments of these analyses differ. An important issue
is the possible incompatibility of different data sets. 
For example, the combination of SLAC and BCDMS data yields  an
\amz~value near to 0.114. Yet, this is known to result
from a superposition of the BCDMS data favouring a value
of about 0.110 with the SLAC data preferring  \as $ \simeq 0.120$. 
Furthermore, the moment analyses, while theoretically advanced to NNLO,
shift the data weight to large $x$ where 
the accuracy of the data is less impressive. Moreover,
there is a dependence of
the result on the minimum $Q^2$ considered~\cite{alfvogt}
which often leads to the introduction of
power corrections with phenomenological $x$ dependence.
Finally the likely presence of $\ln (1/x)$ terms will affect
the data interpretation.
It is thus concluded that the great potential of DIS data to
determine \amz~requires still much more work in order to
determine \as~at the one per cent level of accuracy.

Interesting ideas are pursued to replace 
in the QCD analysis $xg$ by the 
derivative $\partial F_2 / \partial \ln Q^2$~\cite{alfvogt, tung} and
to develop the method of truncated moments~\cite{trunc}
in order to avoid the low $x$ region in analyses 
of structure functions other than $F_2$.
The approach of double asymptotic scaling at low $x$
of $F_2$~\cite{heros}
represents a three parameter solution of the DGLAP 
equations and may lead to a particularly accurate determination 
of \amz~\cite{balfort, uns}.
This solution predicts a steady increase of $xg$
towards low $x$ which yet has to be 
damped at certain $x$ and $Q^2$
since $xg$ may not exceed  the proton size
$\pi r_p^2$ by too big an amount~\cite{levin}. 
\section{On the Future of Deep Inelastic Scattering}
During the year 2000 the HERA luminosity will
be upgraded~\cite{wagner}
in order to provide an integrated luminosity of
150~pb$^{-1}$ per year.
Variations of proton and electron beam energies
and the use of electron polarization in colliding mode will
further enable the electroweak structure function measurements and 
enhance the discovery potential of the machine. The modifications of
HERA are accompanied by  major detector upgrades of the 
luminosity, forward tracking and Silicon vertex detectors
of H1 and ZEUS. 

The main injector neutrino oscillation
detector at Fermilab (MINOS$_{near}$) will lead to
precise, high statistics data ($\simeq 4 \cdot 10^7$
events/year) on the six structure functions
($F_2$, $xF_3$ and $F_L$ for $\nu Fe$ and $\bar{\nu} Fe$ 
scattering) which is necessary to disentangle the nucleon sea, i.e. to
measure $\bu+\bd$, $c$ and $s$ \cite{morfin}.
Measurements of the nuclear dependence
of neutrino DIS cross sections using additional targets
will determine $\nu A$ shadowing and perhaps help resolving
the CCFR-NMC puzzle, Section~\ref{nuexp}.
Increase of neutrino energy by a factor of 10 would be
possible in a 250~GeV muon storage ring providing 
extremely intense neutrino beams~\cite{geer, king}.

The obvious next step in electron-proton DIS is a new $ep$ 
machine~\cite{wwt}.
\begin{figure}[htb]
\begin{center}
\begin{picture}(200,200)
\put(-60,-90){
\epsfig{file=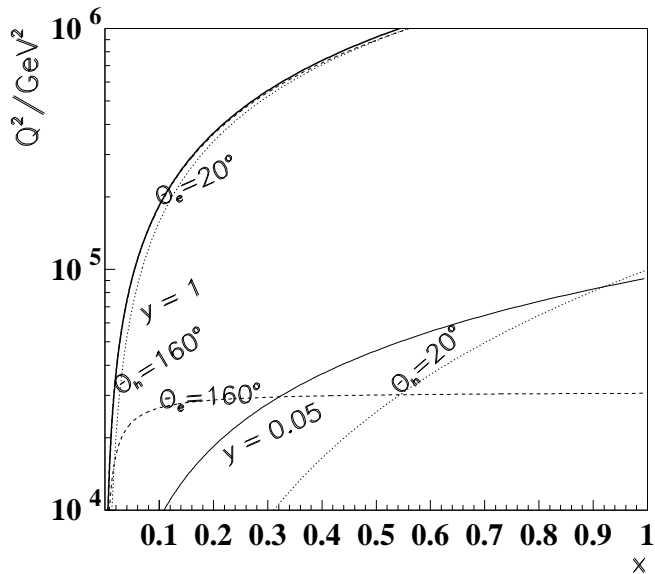,
width=9cm,height=13cm,bbllx=0pt,bblly=0pt,bburx=557pt,bbury=792pt}}
\end{picture}
\caption{Kinematic region of a 
possible future $ep$ collider using 500~GeV
electrons from TESLA and 920~GeV protons from HERA.
The line $y=0.05$ represents the upper kinematic limit of
$ep$ collisions in HERA. Since the
TERA machine is symmetric in energy, it provides full containment of
highly energetic electrons and  hadrons at large $x$ and high $Q^2$
as indicated with the iso-$\theta$ lines at 20$^o$ and 160$^o$.}
\label{tera}
\end{center}
\end{figure}
The proposed linear collider at DESY, TESLA,
 may provide collisions of electrons 
of up to about 500~GeV against HERA protons of nearly
1~TeV. A similar energy of $\sqrt{s} \simeq 1.5~$TeV can be 
obtained in $ep$ collisions at LEP-LHC energies. These machines
differ in technology, luminosity and kinematics. Yet one can 
envisage extending the low $x$ acceptance by a factor of 20 and
DIS data to $Q^2 \simeq 500,000$~GeV$^2$ and beyond, (Fig.~\ref{tera}).
Saturation and sub-structure will be searched for in this extended
range.

 30 years after the pioneering
SLAC $ep$ experiments deep inelastic scattering still has an 
exciting future.
\section{Concluding Remarks}
%
%
HERA has opened the field of low $x$
physics which is governed by gluon interactions
and which is far from being fully understood. The gluon momentum
density at low $x$ is very large. This causes
the structure function $F_2$ to rise at low $x$, it
determines the longitudinal structure function
to be large and the 
production cross section of heavy flavours
to be sizeable.
 Increasing experimental precision leads to sensitive tests
of QCD at higher orders perturbation theory.
Most accurate simultaneous determinations
are in reach of the 
gluon distribution  $and$ the strong interaction
coupling constant with DIS data. Electroweak neutral 
and charged current structure functions provide new insights
in the proton structure at high $x$.
Measurements at $Q^2 \simeq M_Z^2$ probe the proton 
nearly 100 times below 
the parton level reached three decades ago. It is
a spectacular result that no substructure
of leptons or quarks has been observed so far.
At the same time significant progress is made with various 
fixed target and $pp$
experiments leading to deeper insight in the
partonic structure of the proton. The 
gluon distribution at large $x$ is small but remains to be determined.
  The next step is in reach for
tests of the inner proton structure 
down to $2 \cdot 10^{-19}$m. The outcome is unknown
and deep inelastic physics therefore worth continuing effort. \\
\\
\bigskip

It is a pleasure to thank John Jaros and co-organizers for
an excellent Symposium.
Many thanks are due to colleagues from the various DIS experiments
for providing information and guidance in understanding their results.
I have to thank too many individuals to be named here,
 members of the H1 Collaboration and its structure function group,
physicists and engineers of the Zeuthen Silicon detector group,
 colleagues in the DIS99 conference committees,
many theoretical and experimental
physicists for useful discussions and reading the manuscript
and also several known physicists around the BCDMS Collaboration
who introduced me to deep inelastic scattering
years ago. Modern particle physics
is a huge common effort of a large, mostly friendly community.
Particular recognition is due to the youngest: I 
sincerely thank Vladimir Arkadov, Doris Eckstein, Alexander Glazov and
in particular Rainer Wallny for efficient help in preparing this talk
and exciting moments of joint research. 

\end{document}